\documentclass[11pt,oneside]{article}
\usepackage{setspace,graphicx,amssymb,amsmath,latexsym,amsfonts,amscd,amsthm,multirow,ctable,mathdots,caption,array}
\usepackage{fancyhdr,tabularx,cite,mathrsfs}
\usepackage[headings]{fullpage}
\usepackage{stmaryrd}
\usepackage{rotating}
\usepackage{accents}
\usepackage{hyperref}

\newcommand{\bea}{\begin{eqnarray}} 
\newcommand{\eea}{\end{eqnarray}} 
\newcommand{\bee}{\begin{eqnarray*}} 
\newcommand{\eee}{\end{eqnarray*}} 
\newcommand{\al}{\begin{align*}} 
\newcommand{\eal}{\end{align*}} 
\newcommand{\be}{\begin{equation}} 
\newcommand{\ee}{\end{equation}} 
 
\newcommand{\bem}{\begin{pmatrix}} 
\newcommand{\eem}{\end{pmatrix}} 
\newcommand{\thalf}{{\tfrac{1}{2}}}
\newcommand{\half}{{\frac{1}{2}}}
\newcommand{\qrt}{{\tfrac{1}{4}}}
\newcommand{\etabox}[2]{\underset{\ ~#2}{\mbox{\scriptsize $#1$}\ \framebox[15pt]{\phantom{b}}}}
\newcommand{\ubar}[1]{\underaccent{\bar}{#1}}

\newcommand{\nn}{\nonumber}
\newcommand{\lp}{\left(}
\newcommand{\rp}{\right)}

\renewcommand{\d}{\textrm{d}}
\newcommand{\e}{\textrm{e}}

\renewcommand{\a}{\alpha}
\renewcommand{\b}{\beta}

\newcommand{\cZ}{\mathcal{Z}}

\def\a{\alpha} 
\def\b{\beta} 
 
\def\d{\delta} 
 
\def\e{\epsilon}

\def\l{\lambda}

\def\t{\tau} 
\def\th{\theta} 
\def\til{\tilde}

\def\Tr{\tr}

\newcolumntype{R}{ >{$}r <{$}}
\newcolumntype{C}{ >{$}c <{$}}
\newcolumntype{L}{ >{$}l <{$}}
\newcolumntype{F}{>{\centering\arraybackslash}m{1.5cm}}

\newcommand{\mc}[1]{\mathcal{#1}}

\newcommand{\comment}[1]{}
\newcommand{\ba}{\begin{eqnarray}}
\newcommand{\ea}{\end{eqnarray}}
\newcommand{\bb}{\mathbb}

\newcommand{\ZZ}{{\mathbb Z}}%Integers
\newcommand{\tr}{\operatorname{{Tr}}}

\newcommand{\sgn}{\operatorname{sgn}}
\newcommand{\SL}{\operatorname{\textsl{SL}}}      %SL group
\newcommand{\G}{\Gamma}	%Gamma
\newcommand{\g}{\gamma}	%gamma

\newcommand{\rawr}{\includegraphics[height=4.725mm, trim = 0mm 0mm 8cm 0mm, clip]{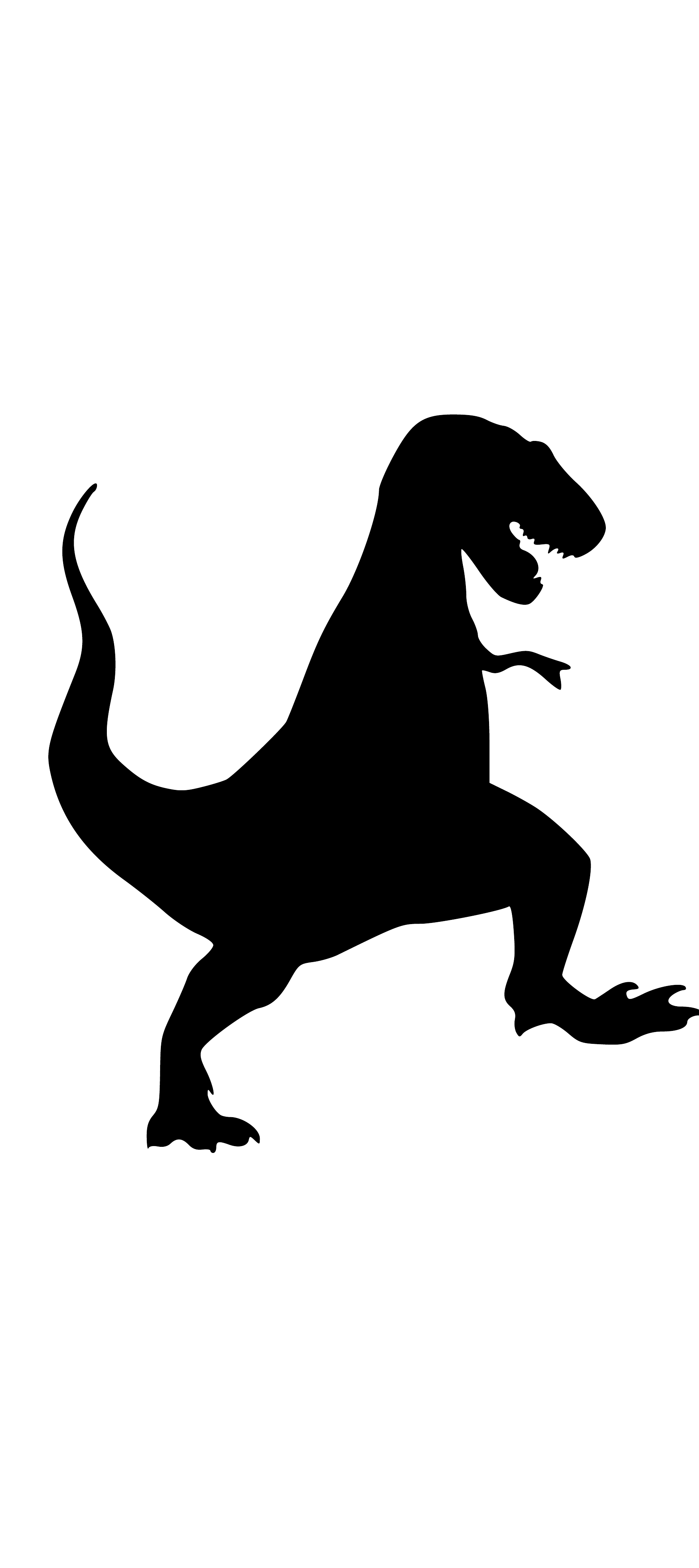}}
\newcommand{\utah}{\includegraphics[height=3.6mm, trim = 1mm -25mm 0mm 0mm, clip]{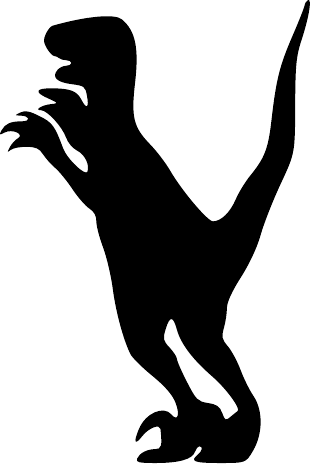}}

\theoremstyle{definition}

\theoremstyle{remark}

\numberwithin{equation}{section}
\def\ket#1{|#1\rangle}

\begin{document}

\setstretch{1.4}
\global\long\def\bbH{\mathbf{H}}
\global\long\def\fullv{X^{i_1}_{n_1} X^{i_2}_{n_2}\ldots X^{i_r}_{n_r}|v\rangle}
\global\long\def\rightv{X^{i_2}_{n_2}\ldots X^{i_r}_{n_r}|v\rangle}

\global\long\def\W{\mathcal{W}}
\global\long\def\lc{\mathopen{:}}
\global\long\def\rc{\mathclose{:}}
\global\long\def\bbC{\mathbf{C}}
\global\long\def\bbZ{\mathbf{Z}}
\global\long\def\bbR{\mathbf{R}}
\global\long\def\SL{\mathrm{SL}_{2}\!\left(\mathbb{Z}\right)}
\global\long\def\calM#1#2{\mathcal{M}_{#1}\!\left(#2\right)}
\global\long\def\sfM#1#2{\mathsf{M}_{#1}\!\left(#2\right)}
\global\long\def\sfS#1#2{\mathsf{S}_{#1}\!\left(#2\right)}
\global\long\def\calJ#1#2#3{\mathcal{J}_{#1}\!\left(#2,#3\right)}
\global\long\def\Ksum{\lim_{K\to\infty}\sum_{\gamma\in\Gamma_\infty\backslash\Gamma_{K,K^2}^*/\Gamma_\infty}}

\global\long\def\sm#1#2#3#4{\left(\begin{smallmatrix}#1  &  #2\cr\cr#3  &  #4\end{smallmatrix}\right)}
\global\long\def\dim{\mathrm{dim}}
\begin{center}

{\LARGE  On the elliptic genera of manifolds of Spin(7) holonomy}
\end{center}
\vspace{0.5cm}
\begin{center}
         {Nathan Benjamin\rawr},
           {Sarah M. Harrison\utah},
                  {Shamit Kachru\rawr},\\
                  {Natalie M. Paquette\rawr},
         and {Daniel Whalen\rawr}

\vspace{0.5cm}\rawr {\it SITP, Department of Physics and
Theory Group, SLAC,\\ Stanford University, Stanford, CA 94305, USA}\\

\vspace{0.5cm}\utah {\it Center for the Fundamental Laws of Nature, Harvard University
 \\ Cambridge, MA 02138, USA}\\

\end{center}

\begin{center}
  {\bf Abstract} 
\end{center}

Superstring compactification on a manifold of Spin(7) holonomy gives rise to a 2d worldsheet conformal field theory with
an extended supersymmetry algebra.  The ${\cal N}=1$ superconformal algebra is extended by
additional generators of spins 2 and 5/2, and instead of just superconformal symmetry one 
has a $c=12$ realization of the symmetry group ${\cal SW}(3/2,2)$.  In this paper, we compute the characters of this supergroup, and decompose the elliptic genus of a general Spin(7) compactification in terms of these characters.  We find suggestive relations to various sporadic groups, which are made more precise in a companion paper.

\newpage
\tableofcontents
\newpage

\section{Introduction}
\label{sec:introduction}

Berger's classification of holonomy groups \cite{Berger} allows for only a few possibilities yielding supersymmetric
vacua of the superstring.  Beyond the low-dimensional avatars of the sequence of spaces of
$SU(n)$ or $Sp(n)$ holonomy (comprising Calabi-Yau threefolds and fourfolds and low-dimensional hyperK\"ahler manifolds), there
are two exceptional holonomy groups that are relevant: G$_2$ and Spin(7).  Compact examples of such spaces, of possible interest for string compactification, were first constructed by Joyce -- for Spin(7) holonomy good discussions appear in \cite{joyce1996compact, Joyce}.  Our goal in this paper is
to compute the elliptic genera of Spin(7) manifolds, and to make some observations 
about interesting connections between geometry, number theory, and group theory at which they hint.  A more precise version of these
connections is derived in the companion paper \cite{M5spin7}.

Our study will proceed as follows.  In \S\ref{sec:genera}, we derive on macroscopic grounds (using simple arguments
about the allowed space of modular functions) the elliptic genus of a Spin(7) manifold $X$.  This yields
a one-parameter family of modular functions under the congruence subgroup $\Gamma_{\theta}$;
the single parameter is determined by the Euler character $\chi(X)$.

We then wish to decompose the elliptic genus into a sum of irreducible characters of the worldsheet
superconformal field theory.  It has been known since the work of Shatashvili and Vafa \cite{shatashvili1995superstrings} that the ${\cal N}=1$ superconformal algebra is extended in
this case by the addition of new generators of spins 2 and 5/2.  This yields a $c=12$ realization
of the ${\cal SW}(3/2,2)$ algebra.  While many properties of the representations
were studied in \cite{gepner2001unitary}, the full set of characters have not yet been determined (although a conjecture for the massive characters, which we will confirm, appears in \cite{eguchi2003supercoset}).  In \S\ref{sec:charactersspin7} we discuss the derivation of the characters of the algebra.   

In \S\ref{sec:decomps}, we decompose the elliptic genus into irreducible characters.  We find a suggestive connection with representation theory of various sporadic simple groups. 
It is difficult, however, to make this connection precise beyond the level of numerology for reasons
we discuss in \S\ref{sec:conclusions}.
A precise connection between a
chiral conformal field theory with the same superconformal symmetry group and these groups
appears in \cite{M5spin7}.

Indeed, a good part of our motivation for undertaking this study was the 
2010 observation of Eguchi, Ooguri, and Tachikawa (EOT) regarding K3 models \cite{EOT}.   K3 sigma models
enjoy ${\cal N}=(4,4)$ superconformal symmetry.  EOT
observed that in the decomposition of the elliptic genus of K3 in terms of $\mathcal{N}=4$ characters, the coefficients could be expressed as sums of dimensions of irreducible representations of M$_{24}$, the largest Mathieu group. Explicitly, the characters of the $\mathcal{N}=4$ algebra are
\ba
\operatorname{ch}_{h=\frac14,l=0}(\t,z) &=& -\frac{i y^{\frac12} \theta_1(\t,z)}{\eta(\t)^3} \sum_{n=-\infty}^{\infty} \frac{(-1)^n q^{\frac12 n(n+1)} y^n}{1-y \ q^n}\\ 
\operatorname{ch}_{h=n+\frac14,l=\frac12}(\t,z) &=& q^{n-\frac18}\frac{\theta_1(\t,z)^2}{\eta(\t)^3}
\ea
in terms of which the K3 elliptic genus may be written
\ba
\cZ_{elliptic}^{K3}(\t,z) &=& 8 \left[ \lp \frac{\theta_2(\t,z)}{\theta_2(\t,0)}\rp^2 + \lp \frac{\theta_3(\t,z)}{\theta_3(\t,0)}\rp^2 + \lp \frac{\theta_4(\t,z)}{\theta_4(\t,0)}\rp^2 \right]\\
&=& 24 \operatorname{ch}_{h=\frac14,l=0}(\t,z) + \sum_{n=0}^\infty A_n \operatorname{ch}_{h=n+\frac14,l=\frac12}(\t,z)\,.
\ea
The observation of EOT, applied to the first few coefficients, is:
\ba\label{eq:As}
A_0 &=& -2=-1-1\,,\nn\\
A_1 &=& 90 = 45 + \overline{45}\,,\nn\\
A_2 &=& 462 = 231 + \overline{231}\,,\nn\\
A_3 &=& 1540 = 770 + \overline{770}\,,\nn\\
A_4 &=& 4554 = 2277+2277\,,\nn\\
&\ldots& 
\ea

The decompositions and twining genera were put on firm footing in \cite{Miranda, Gaberdiel:2010ch, Gaberdiel:2010ca, Eguchi:2010fg}.
It was proven in \cite{Gannonproof} that all the $A_n$ for $n\geq1$ are sums of irreducible representations of M$_{24}$ with only positive coefficients. However, a natural construction of the full M$_{24}$ group acting on a module, such as one associated to K3 sigma models, is unknown.  Mukai \cite{Mukai} showed that the groups of symplectic automorphisms of any K3---that is, automorphisms that act trivially on the holomorphic 2-form---fall into certain subgroups of M$_{23}$ and hence are insufficient to explain the appearance of M$_{24}$. Moreover, \cite{Gaberdiel} showed that no K3 sigma model admits the full M$_{24}$ as a symmetry and some sigma models possess symmetries that lie outside M$_{24}$, yet in Co$_1$.  Further aspects of this moonshine and its precise
relation to K3, as well as various extensions, have been explored in \cite{Cheng-Duncan,Cheng:2012tq,Cheng:2013kpa,Cheng:2013wca,Cheng:2014owa,Harrison:2013bya,Cheng:2014zpa,Taormina:2011rr,Taormina:2013jza,Taormina:2013mda,Gaberdiel:2013psa,Harvey:2013mda,paquette2014comments,Harvey:2014cva}. For a recent review, see \cite{DuncanOno}.

While the precise nature of the relation of M$_{24}$ to K3 $\sigma$-models remains unclear,
it is clearly worthwhile to look for similar connections in other classes of supersymmetric string
compactifications.  Especially relevant to our work are various precise
versions of mock modular moonshine associated with the superconformal field theory on the chiral
$E_8$ lattice \cite{Cheng:2014owa} (building on the earlier work in \cite{JohnConway}; see also the recent related work \cite{DuncanMackCrane}).  Our study of Spin(7) manifolds here suggested an extension of
the latter approach, and indeed using the same characters we find here, a precise moonshine
conjecture relating M$_{24}$, a certain $c=12$ conformal field theory with ${\cal SW}(3/2,2)$
symmetry, and various mock modular forms appears in the closely related companion work \cite{M5spin7}.  

Readers primarily interested in the main results of the paper can skip directly to \S\ref{sec:decomps}.  

\section{Elliptic Genus for Spin(7) manifolds}\label{sec:genera}

Let us begin by calculating the elliptic genera for manifolds of Spin(7) holonomy. We will derive the answer using general arguments based on modularity and pole structure, and then check our result using orbifold techniques to directly compute the answer in some of the examples constructed by Joyce \cite{Joyce, joyce1996compact}. 

\subsection{Genera(lities)}
The elliptic genus can be defined for any compact even-dimensional spin manifold\footnote{The genus as defined here vanishes for odd-dimensional manifolds.} \cite{hirzebruch1992manifolds}. Let's first consider the NS$,+$ elliptic genus, defined as
\be
Z_{NS,+}(\t) = \Tr_{NS,R} (-1)^{F_R} q^{L_0-c/24}\bar{q}^{\bar{L}_0 - c/24}.
\label{eq:egfull}
\ee
Note that this trace acts as a Witten index on the right-movers, so the $\bar{q}$ dependence drops out leaving us with a holomorphic object.

It will also be convenient to define elliptic genera with different boundary conditions and insertions in the trace:
\begin{align}
\chi &= \Tr_{R,R} (-1)^F q^{L_0-c/24}\bar{q}^{\bar{L}_0 - c/24} \nn \\
Z_{R,+}(\tau) &= \Tr_{R,R} (-1)^{F_R} q^{L_0-c/24}\bar{q}^{\bar{L}_0 - c/24} \nn \\
Z_{NS,-}(\tau) &= \Tr_{NS,R} (-1)^F q^{L_0-c/24}\bar{q}^{\bar{L}_0 - c/24} 
\end{align}
where $(-1)^F$ is the fermion sign operator. $\chi$ is the Witten index, and the rest of the genera are functions of $\tau$, which are invariant under a level 2 congruence subgroup of $SL(2,\bb{Z})$ that preserves the appropriate spin structure on the torus.

By considering elements of $SL(2,\bb{Z})$ that exchange spin structures, we can relate $Z_{R,+}$, $Z_{NS,-}$, and $Z_{NS,+}$ to each other. In particular
\begin{align}\label{eq:relations}
Z_{R,+}(\tau) &= Z_{NS,-}(-1/\tau) \nn \\
Z_{R,+}(\tau) &= -Z_{NS,+}(-1/\tau + 1).
\end{align}

The congruence subgroup that preserves the spin structure of $Z_{NS,+}(\tau)$ (antiperiodic fermions in space and time) is given by
\be\Gamma_\theta=\left\{\gamma\in SL(2,\bb{Z})\;\middle|\; \gamma\equiv\sm1001\text{ or }\sm0110\bmod 2 \right\},\ee
which is the subgroup generated by $\tau \rightarrow \tau+2$ and $\tau \rightarrow -1/\tau$. The fundamental domain of $\G_{\th}$ has genus 0, and the corresponding normalized (i.e. with constant removed) Hauptmodul is
\begin{align}
K(\tau) &= \dfrac{\Delta(\tau)^2}{\Delta(2\tau) \Delta(\tau/2)}- 24 \nn \\
&= q^{-1/2} + 276 q^{1/2} + 2048 q + O(q)^{3/2},
\end{align}
where $\Delta(\tau) = \eta(\tau)^{24}$ and $q=\exp(2\pi i\tau)$. In particular, any function invariant under $\G_{\th}$ that is meromorphic on the upper half-plane and at $\infty$ can be written as a rational function of $K(\tau)$. By looking at the pole structure of a function we can specify the rational coefficients of $K(\tau)$.

Just as in \cite{witten2007three}, the function $Z_{NS,+}$ is convergent for $0 < |q| < 1$ due to the $\Tr q^{L_0}$ so there are only potential poles at $\t = i \infty$ and $\t = 1$. The NS ground state energy of $-\tfrac{d}{16}$ gives $Z_{NS,+}(\tau)$ a pole of order $\tfrac{d}{16}$ at $\tau=i\infty$, where $d$ is the real dimension of the manifold. For Spin(7) manifolds, $d=8$ and the pole will be of order $\frac{1}{2}$. At $\tau=1$, the other cusp, the function is regular because the Ramond sector genus $Z_{R, +}$ is regular at the high-temperature cusp (since its ground states have zero energy). Transforming to $Z_{NS, +}$ using (\ref{eq:relations}) we see $$\lim_{\tau \rightarrow i \infty} Z_{R, +}(\tau) = - Z_{NS, +}(1).$$
Thus we can write the elliptic genus as
\be
Z_{NS,+}(\tau) = c_0 + c_1 K(\tau)
\label{eq:egk}
\ee
for constants $c_0$ and $c_1$. From (\ref{eq:relations}), we can get the R+ elliptic genus as well.

\subsection{Specialization to Spin(7)}

The only remaining freedom is to fix the two coefficients in (\ref{eq:egk}).  To do this, we now turn to geometric properties of the elliptic genus.  (Alternatively, one could try to match the results to explicit computations in two models, particularly if one knows which topological quantities the constants depend on.)

Witten emphasized the geometric interpretation of the elliptic genus as a character-valued index of a (twisted) Dirac operator on loop space $\mathcal{L}M$, for any spin manifold $M$ with tangent bundle\footnote{One could of course consider more general vector bundles on the manifold, but we will not do this here.} $T_M$~\cite{witten1988index}. Construct the following symmetric and antisymmetric products of bundles:
\be \label{eq:bundle}
R(T_M) = \bigotimes_{\substack{l \in \mathbb{Z}\\l>0}} S_{q^l}(T_M) \otimes \bigotimes_{\substack{l \in \mathbb{Z} + \frac12\\l>0}} \bigwedge_{q^l}(T_M)
\ee
where we have defined $\bigwedge_t T_M = \sum_{i=0}^{dim M} \bigwedge^i T_M t^i$ and $S_t T_M = \sum_{i=0}^{\infty} S^i T_M t^i$. Then the NS-R elliptic genus for the even-dimensional manifold $M$ with ${\rm dim} M= d$ can be written as 
\be \label{eq:geomgenus}
Z_{NS, +} = q^{-d/16} \left\langle \hat{A}(M)\operatorname{ch} R(T_M), M \right\rangle~.
\ee
The inner product is often rewritten as an integral over $M$.  Here, $\hat{A}(M)$ is the $\hat{A}$-class of the manifold. If M has total Pontryagin class $p(T_M)= \prod_{i}(1 + u_i)$ then we may write $\hat{A}(M) = \prod_{i}\frac{\sqrt{u_i}/2}{\sinh(\sqrt{u_i})/2}$, where we have used the splitting principle. In what follows, we will frequently use the relation between Pontryagin and Chern classes: $p_{i}(T_M) = (-1)^i c_{2 i}(T_{\mathbb{C}})$.

The Fourier development of the elliptic genus is given by
\be
Z_{NS, +} = q^{-d/16} \sum_{l}\textrm{index}(R_l)q^{l/2}
\ee
where $R_{l}$ is the representation of $R(T_M)$ multiplying the $l^{\text{th}}$ power of $q$ in (\ref{eq:bundle}). For Spin(7) manifolds we want to match this to equation (\ref{eq:egk}); to fix two undetermined constants, we need to determine two of the $R_l$.  We will choose
the simplest cases, $R_0 = 1$ and $R_1 = T_{M}$, and compute their indices below.

A comment before we get started: the $\hat{A}$-genus, which is the evaluation of the above $\hat{A}$-class $\hat{A}(M)$ on the fundamental class\footnote{In plain language, this is an instruction to take the form of top degree in the expansion of $\hat{A}(M)$ and integrate over M.} of the manifold, is itself the index of the Dirac operator\footnote{In the K\"{a}hler case, this is equivalent to the index of $\bar{\partial}$ acting on $(0, q)$ forms, which becomes the holomorphic Euler characteristic $\chi_0 = \sum_{q} (-1)^q H^{0, q}(M, \mathbb{Z})$.}. Joyce proved that for simply-connected Spin(7) manifolds, this quantity is simply equal to 1 \cite{joyce1996compact}. This fact follows from the formula for the $\hat{A}$-genus, $24 \hat{A}  = -1 + b^1 - b^2 + b^3 + b^4_{+} - 2 b^4_{-}$, and a constraint on the Betti numbers of Spin(7) manifolds: $b_3 + b_4^{+}- b_2 - 2 b_4^{-}-1= 24$. Note that the formula for the $\hat{A}$-genus can be rewritten in terms of Pontryagin classes as $5760 \hat{A} = 7 p_1^2 - 4 p_2$. In the previous formula and below we will suppress the argument of the Pontryagin classes and so write $p_i$ to denote $p_i(T_M)$. We will also abuse notation by using the same expression before and after integrating over M.

First we look at the $q^{-1/2}$ coefficient. We have ${\rm index} (R_0) =\int_{M} \hat{A}(M) \operatorname{ch}(1)$. For an 8-manifold, we can expand $\hat{A}(M) = 1 - \frac{p_2}{24} + \frac{7 p_1^2 - 4 p_2}{5760}$. Integrating over $M$ and using Joyce's result that the $\hat{A}$-genus equals 1 we just get a coefficient of 1. (In the Calabi-Yau case we would get $\hat{A}= 2$ because such manifolds preserve more supersymmetry.) In other words, we have just counted the dimension of the space of harmonic spinors. 

We also want to find the $q^0$ coefficient, which is given by ${\rm index}(T)= \int_M \hat{A}(M) \operatorname{ch}(T)$. For any complex bundle $V$ on an 8-manifold we can expand the Chern character as $\operatorname{ch}(V) = {\rm dim(V)} + c_1(V) + \thalf(c_1(V)^2 - c_2(V)) + \frac{1}{3!}(c_1(V)^3 - 3 c_2(V) c_1 (V) + 3 c_3(V)) + \frac{1}{4!}(c_1(V)^4 - 4 c_2(V) c_1(V)^2 + 4 c_3(V) c_1(V) + 2 c_2(V)^2 - 4 c_4(V))$. For us, $\operatorname{ch}(T)$ denotes the Chern character of the complexification of the tangent bundle, $T_{\mathbb{C}} = T \oplus i T$, though below we will suppress this argument as well. Setting $c_1=0$, multiplying by $\hat{A}(M)$, and integrating we are left with
\be 
{\rm index}(T) = {\rm dim}(T) \frac{7 p_1^2 - 4 p_2}{5760} + \frac{p_1 c_2}{24} + \frac{1}{4!}(2 c_2^2 - 4 c_4) 
\ee
where we can set the first term to ${\rm dim} (T)$ because it is multiplied by the $\hat{A}$-genus, which is just~1. It is convenient to use the relation $4 p_2 - p_1^2 = 8 \chi$ (see the remark below the proof of 7.3 in \cite{salamon}). Employing the relationship between Chern and Pontryagin classes above, $c_4 = \chi = p_2$ and $-c_2 = p_1$, the relation from \cite{salamon} becomes $c_2^2 = - 4 \chi$.

Then we have
\be 
{\rm index}(T) = {\rm dim}(T) - \frac{c_2^2}{24} + \frac{1}{4!}(2 c_2^2 - 4 \chi) = -\frac{\chi}{6} + \frac{c_2^2}{24} + {\rm dim}(T) = -\frac{\chi}{3} + 8.
\ee
In the last line we have used ${\rm dim}(T) = {\rm dim}_{\mathbb{C}} (T_{\mathbb{C}}) ={\rm  dim}_{\mathbb{R}}(T)$. Our final result, then, is that for a Spin(7) manifold $M$, one has
\begin{equation}
\label{theanswer}
Z_{NS, +}(M) = K(\tau) + \left( 8 - {\frac{\chi}3} \right)~=~q^{-1/2} + \left( 8 - {\frac{\chi}3}\right)
+ 276 q^{1/2} + 2048 q + \ldots
\end{equation}
In the rest of the paper, we will decompose (\ref{theanswer}) in terms of characters of
${\cal SW}(3/2,2)$, and observe some interesting structure in the coefficients.  We discuss the
relevant numerology in \S\ref{sec:decomps}.

\subsection{Some explicit computations}

We now compute a few examples of Spin(7) elliptic genera directly, using orbifold techniques. We will see that they obey the general formula (\ref{theanswer}). In \cite{Joyce}, Joyce constructs several examples of manifolds with Spin(7) holonomy. We will focus on his toroidal orbifolds of the form $T^8/\bb{Z}_2^4$.  Our computations follow the techniques used in \cite{David:2006ji}.  For our conventions regarding theta functions, see Appendix \ref{app:a}.

Joyce gives three examples of $T^8/\bb{Z}_2^4$ orbifolds with Spin(7) holonomy. 
Label the torus coordinates by $(x_1, \ldots, x_8)$.  The orbifolds are as follows.

\begin{enumerate}
\item The first example Joyce gives has the following orbifold action: 
\begin{align}
\alpha: (x_1, \ldots, x_8) &\mapsto (-x_1, -x_2, -x_3, -x_4, x_5, x_6, x_7, x_8) \nonumber \\
\beta: (x_1, \ldots, x_8) &\mapsto (x_1, x_2, x_3, x_4, -x_5, -x_6, -x_7, -x_8) \nonumber \\
\gamma : (x_1, \ldots, x_8) &\mapsto (\thalf-x_1, \thalf-x_2, x_3, x_4, \thalf-x_5, \thalf-x_6, x_7, x_8) \nonumber \\
\delta: (x_1, \ldots, x_8) &\mapsto (-x_1, x_2, \thalf-x_3, x_4, -x_5, x_6, \thalf-x_7, x_8).
\end{align}

We need to sum over the untwisted sector and all fifteen twisted sectors, each of which must be projected onto $\bb{Z}_2^4$-invariant states by inserting the product
$\left(\frac{1+\a}{2}\right) \left(\frac{1+\b}{2}\right) \left(\frac{1+\g}{2}\right)\left(\frac{1+\d}{2}\right)$
into the trace. In total we will need to sum over 256 separate boundary conditions. Fortunately, most of the sectors' contributions to the elliptic genus will vanish.

Any sector with both an untwisted space and an untwisted time boundary condition along any of the eight coordinates will have a right-moving fermion zero-mode and thus its contribution will vanish. Similarly, any sector with both a twisted space and a twisted time boundary condition along any of the eight coordinates will have a left-moving fermion zero-mode and its contribution also vanishes. Finally, any insertions in the trace that permute fixed points will not contribute.

The surviving twists are
\be
Z_{NS,+} = 16 \left( \etabox{1}{\a\b} + \etabox{\a\b}{\phantom{al} 1} + \etabox{\b} \a + \etabox{\a} \b \right)
\ee
where the 16 includes both the right-moving Ramond ground states, and the fixed point contributions.

Calculating the contributions of each twisted sector to the elliptic genus is an exercise in free field theory. For example, the $\etabox{1}{\a\b}$ contribution involves a spatial twist for all eight bosons and fermions, and no time twist, making the bosons antiperiodic in space, and the fermions periodic in space (recall we have NS boundary conditions). We insert nothing in the trace. Finally we have to take into account the ground state energy of $q^\half$ and ground state degeneracy. The contribution to the elliptic genus is 
\begin{align}
\etabox{1}{\a\b} &= 4 q^\thalf \prod_{n=1}^{\infty} \left( \frac{1+q^n}{1-q^{n-\thalf}} \right)^8 = \frac{\th_2(\t,0)^4}{\th_4(\t,0)^4}.
\end{align}
The sum over all sectors is
\begin{align}
Z_{NS,+} &= 16 \left(\frac{\th_2(\t,0)^4}{\th_4(\t,0)^4} + \frac{\th_4(\t,0)^4}{\th_2(\t,0)^4} - 2 \right) \nn \\
&= \frac{1}{\sqrt{q}} - 40 + 276 \sqrt{q} + 2048 q + 11202 q^{3/2} + \ldots \nn \\
&= K(\tau) - 40.
\end{align}
If we do a similar computation for the R+ elliptic genus, we get
\begin{align}
Z_{R,+} &= 16 \left(\frac{\th_3(\t,0)^4}{\th_4(\t,0)^4} + \frac{\th_4(\t,0)^4}{\th_3(\t,0)^4} + 2 \right) \nn \\
&= 64 + 4096q + 98304q^2 + 1228800q^3 + \ldots
\end{align}

Joyce gives a resolution of this orbifold to a smooth manifold with Euler character $\chi = 144$, which matches (\ref{theanswer})\cite{joyce1996compact, Joyce}.

\item The second example is a $T^8/\bb{Z}_2^4$ with the $\bb{Z}_2$'s acting as
\begin{align}
\a: (x_1, \ldots, x_8) &\mapsto (-x_1, -x_2, -x_3, -x_4, x_5, x_6, x_7, x_8) \nonumber \\
\b: (x_1, \ldots, x_8) &\mapsto (x_1, x_2, x_3, x_4, -x_5, -x_6, -x_7, -x_8) \nonumber \\
\g: (x_1, \ldots, x_8) &\mapsto (\thalf-x_1, \thalf-x_2, x_3, x_4, -x_5, -x_6, x_7, x_8) \nonumber \\
\d: (x_1, \ldots, x_8) &\mapsto (-x_1, x_2, -x_3, x_4, \thalf-x_5, x_6, \thalf-x_7, x_8).
\end{align}

The contributing twists to the elliptic genus in this example are the same as in the previous, so we again get
\begin{align}
Z_{NS,+} &= 16 \left(\frac{\th_2(\t,0)^4}{\th_4(\t,0)^4} + \frac{\th_4(\t,0)^4}{\th_2(\t,0)^4} - 2 \right) \nonumber \\
&= \frac{1}{\sqrt{q}} - 40 + 276 \sqrt{q} + 2048 q + 11202 q^{3/2} + \ldots 
\end{align}
and similarly
\begin{align}
Z_{R,+} &= 16 \left(\frac{\th_3(\t,0)^4}{\th_4(\t,0)^4} + \frac{\th_4(\t,0)^4}{\th_3(\t,0)^4} + 2 \right) \nn \\
&= 64 + 4096q + 98304q^2 + 1228800q^3 + \ldots
\end{align}

Joyce gives several inequivalent resolutions of this orbifold, but all of them have Euler character $\chi = 144$.

\item Joyce provides a final example with $\bb{Z}_2$ group actions
\begin{align}
\a: (x_1, \ldots, x_8) &\mapsto (-x_1, -x_2, -x_3, -x_4, x_5, x_6, x_7, x_8) \nonumber \\
\b: (x_1, \ldots, x_8) &\mapsto (x_1, x_2, x_3, x_4, -x_5, -x_6, -x_7, -x_8) \nonumber \\
\g: (x_1, \ldots, x_8) &\mapsto (-x_1, -x_2, x_3, x_4, -x_5, -x_6, x_7, x_8) \nonumber \\
\d: (x_1, \ldots, x_8) &\mapsto (\thalf-x_1, x_2, \thalf-x_3, x_4, \thalf-x_5, x_6, \thalf-x_7, x_8).
\end{align}

Unlike in the previous two examples, this orbifold has three $\bb{Z}_2$ actions that do not involve coordinate shifts whereas the previous examples had only two. Thus we get more sectors contributing to the elliptic genus. The contributing twists are
\begin{align}
Z_{NS,+} &= 16 \left( \etabox{1}{\a\b} + \etabox{\a\b}{\phantom{al}1} + \etabox{\b} \a + \etabox{\a} \b + \etabox{\b\g}{\phantom{a}\a\g} + \etabox{\a\g}{\phantom{a}\b\g} + \etabox{\g}{\a\b\g} + \etabox{\a\b\g}{\phantom{aal} \g} \right) \nonumber \\
&= 16 \left( \frac{\th_2(\t,0)^4}{\th_4(\t,0)^4} + \frac{\th_4(\t,0)^4}{\th_2(\t,0)^4} - 6 \right) \nonumber \\
&= \frac{1}{\sqrt{q}} - 104 + 276 \sqrt{q} + 2048 q + 11202 q^{3/2} + \ldots
\label{eq:ex3}
\end{align}
The same analysis in the R+ elliptic genus gives
\begin{align}
Z_{R,+} &= 16 \left(\frac{\th_3(\t,0)^4}{\th_4(\t,0)^4} + \frac{\th_4(\t,0)^4}{\th_3(\t,0)^4} + 6 \right) \nn \\
&= 128 + 4096q + 98304q^2 + 1228800q^3 + \ldots
\end{align}

Joyce gives several inequivalent resolutions of this orbifold, and it has Euler character $\chi = 336 - 48k - 48l + 12kl$, with $k, l = 0, 1, \ldots 8$.  The orbifold with these choices of phases (as opposed
to possible other choices of discrete torsion) matches onto the results Joyce finds for resolutions with
$k=l=0$ or $k=l=8$.

\item We can take the same toroidal orbifold as above and put in some choice of discrete torsion: in each twisted sector Hilbert space $\mc{H}_h$, introduce a phase $\e(g, h)$, as done in \cite{Vafa:1986wx}. In fact, at each fixed point, we can introduce a different choice of torsion, $\e_f(g,h)$, as done in \cite{MatthiasG2}. However, there are consistency conditions that must be satisfied. In particular, in order for the action to be a representation of $\bb{Z}_2^4$, we require 
\be
\e_f(g,h) \e_f(g',h) = \e_f(gg',h).
\label{eq:constraint1}
\ee
Moreover, modular invariance requires
\begin{align}
\etabox gh(\t+1) &= \etabox{gh}h(\t) \nn \\ 
\etabox gh(-1/\t) &= \etabox{h^{-1}}g(\t).
\label{eq:constraint2}
\end{align}

Note that for $\bb{Z}_2^4$, (\ref{eq:constraint1}) and (\ref{eq:constraint2}) imply $\e_f(g,1) = \e_f(1,g) = 1$ for all $g$, and that all the phases $\e_f(g,h)$ are $\pm 1$ since $\e_f(g,h)^2 = \e_f(1,h) = 1.$

We can rewrite the first line of (\ref{eq:ex3}) in full generality as 
\begin{align}
Z_{NS,+} &= 16\left(\etabox{1}{\a\b} + \etabox{\a\b}{\phantom{al}1}\right) + \sum_f \left(\e_f(\b, \a)\etabox{\b} \a \right)+ \sum_f \left(\e_f(\a, \b) \etabox{\a} \b\right)
\nn\\&+ \sum_f \left(\e_f(\b\g, \a\g) \etabox{\b\g}{\phantom{a}\a\g}\right)+ \sum_f\left(\e_f(\a\g, \b\g) \etabox{\a\g}{\phantom{a}\b\g}\right)\nn\\
&+ \sum_f\left(\e_f(\g, \a\b\g) \etabox{\g}{\a\b\g}\right) + \sum_f \left(\e_f(\a\b\g, \g) \etabox{\a\b\g}{\phantom{aal} \g}\right).
\label{eq:premodstuff}
\end{align}

Using (\ref{eq:constraint2}) repeatedly, we can rewrite every term in (\ref{eq:premodstuff}) terms of the $\a\b$-twisted sector: $\e_f(g, \a\b)$ for some $g$. Then we need to make a consistent choice of torsion for each of the fixed points in the $\a\b$-twisted sector in order to solve for the elliptic genus. The choice 
\begin{align}
\e_f(\a, \a\b) &= -1 \nn \\
\e_f(\g, \a\b) &= -1
\end{align}
for all $128$ fixed points $f$ is a consistent choice which gives 
\begin{align}
Z_{NS,+} &= 16\left( \frac{\theta_2(\t,0)^4}{\theta_4(\t,0)^4} + \frac{\theta_2(\t,0)^4}{\theta_4(\t,0)^4} + 2 \right) \nn\\
&=\frac{1}{\sqrt{q}} + 24 + 276 \sqrt{q} + 2048 q + 11202 q^{3/2} + \ldots
\end{align}
which corresponds to a manifold with $\chi = -48$. Joyce's resolutions with $k=0$, $l=8$ or $k=8$, $l=0$ give such a $\chi$.

The same choice of torsion in the R+ elliptic genus gives
\begin{align}
Z_{R,+} &= 16 \left(\frac{\th_3(\t,0)^4}{\th_4(\t,0)^4} + \frac{\th_4(\t,0)^4}{\th_3(\t,0)^4} - 2 \right) \nn \\
&= 4096q + 98304q^2 + 1228800q^3 + \ldots
\end{align}

It might be interesting to match possible choices of torsion in this class of
models with all possible geometric resolutions; similar results in the context of manifolds of G$_2$
holonomy were obtained in \cite{MatthiasG2}.

\end{enumerate}

\section{Towards Characters of the ``Spin(7) Algebra"}
\label{sec:charactersspin7}

In this section, we review the structure of the chiral algebra for a sigma-model with target a manifold of Spin(7) holonomy. 
The algebra was first studied in the context of Spin(7) compactifications in \cite{shatashvili1995superstrings}.  We discuss our conventions and present explicit details in Appendix~\ref{ap:comm}. 

The reduced holonomy of a Spin(7) manifold $M$ implies the existence of a nowhere-vanishing self-dual 4-form, $\Omega$. $\Omega$ can be written in closed form in terms of a local vielbein $\sum_{i=1}^{8} e_i \otimes e_i$ (see equations $2.1-2.3$ of \cite{shatashvili1995superstrings}), where the $e_i$ are in the fundamental of $O(8)$.  We may choose an embedding $\text{Spin(7)} \subset O(8)$ such that the 8-dimensional spinor representation of $O(8)$ decomposes as $\textbf{7}\oplus \textbf{1}$. Then viewing the 8-dimensional vector representation of $O(8)$ as a spinor of Spin(7), the fourfold antisymmetric product of this spinor includes the singlet $\Omega$. Moreover, a Spin(7) manifold is characterized by three independent Betti numbers that we can express as $b_2, b_3, b_4^{\pm}$ subject to the constraint $b_3 + b_4^{+}- b_2 - 2 b_4^{-}-1= 24$. The dimension of the Spin(7) moduli space is $b_4^{-} + 1$. 

In order to find a geometry-inspired construction of the algebra, we start with the $\mathcal{N}=1$ superconformal algebra (SCA) and extend it by new generators. In the case of a Calabi-Yau sigma model, one must add a $U(1)$ current and impose closure of the algebra. In the case of a Spin(7) manifold, something more exotic happens. The analog of the $U(1) = U(n)/SU(n)$ is in this case a sector isomorphic to $SO(8)/\text{Spin(7)}$. Computing the central charge of this coset, $4 - 7/2 = 1/2$, we can see that the result is the Ising model. Thus, the Ising sector will also produce an analogue of the spectral flow isomorphism enjoyed by Calabi-Yau sigma models. 

As worked out in detail by Shatashvili and Vafa using a free-field representation, the algebra is generated by the operators:
\be 
{L_n, G_n, X_n, M_n}
\ee
where $L_n, G_n$ are the usual Virasoro generators and their superpartners. $X_n$ are the Fourier modes of a new spin-2 operator produced by replacing the $e_i$ in $\Omega$ with target space fermions. Checking the OPEs and enforcing closure of the algebra results in a spin-5/2 superpartner for $X_n$ that we call $M_n$. Note that $G_n, M_n$ are fermionic and will both be half-integrally graded in the NS-sector. 
One can check that $X_n$ is related to an Ising model stress-energy tensor $T_I$ by $T_I = X/8$; see Appendix~\ref{ap:comm} for the full algebra.

Finally, we note in passing that even in the absence of a $U(1)$ current there is a version of ``spectral flow" enabling one to transform between the NS and R sectors. Consider the R-sector ground states, which have $h= c/24 = 1/2$. Label each state by $|h_{R}, h_I \rangle$ where $h_I$ is the eigenvalue of $T_I$ and $h_R$ is the eigenvalue of $T_R$ such that $T= T_R + T_I$. For a unitary theory we have only three classes of such states, because there are only three admissible weights in the Ising sector: $0, \frac1{16}, \frac12$. Therefore, one of these ground states has weight in the Ising sector only: $h_{tot}= h_{I}= 1/2$. Shatashvili and Vafa utilize the fusion rules for this operator, which is isomorphic to the Ising energy operator $\epsilon$, to map Ramond ground states to certain NS sector highest weight states and show it generates the spectral flow-like isomorphism. We will use this isomorphism to map our NS-sector characters to the R-sector and thus obtain the complete set of characters.

\subsection{Unitary Highest Weight States}
Next, we would like to study the representation theory of this algebra and locate the unitary highest weight states in the NS sector, in order to compute their characters. Here, we will briefly sketch some results of Gepner and Noyvert \cite{gepner2001unitary}, who first extensively studied the representation theory of the algebra and computed the Kac determinant\footnote{The operators in \cite{shatashvili1995superstrings} and \cite{gepner2001unitary} are related by: $X(z)= 8 A(z), \ M(z) = \frac{\sqrt{23}}{3} U(z) + \frac{1}{6} \partial G(z)$.}. The Kac determinant is a useful first step in finding the irreducible representations of the algebra and for many algebras, like the Virasoro algebra, it is also sufficient.

We will see, however, that the Kac determinant fails to provide complete information about the structure of the maximal proper submodule generated by null states in several important ways.

Gepner and Noyvert evaluated the Kac determinant for the $\mathcal{SW}(3/2, 2)$ highest weight modules using the Coulomb gas formalism.  Highest weight vectors are written as $|h,x\rangle$ and satisfy
\begin{align} 
X_0| h, x \rangle = x | h, x \rangle \nonumber\\ 
L_0| h, x \rangle = h | h, x \rangle \nonumber\\ 
\mathcal{O}_n | h, x \rangle = 0, \forall n >0. 
\end{align}
The Kac determinant of such a module has a closed-form expression 
\be 
det M^{NS}_{N}(h, a) = \prod_{1 \leq mn \leq 2N} (f_{m, n})^{P_{NS}(N - mn/2)} \prod_{1 \leq jk \leq N}(g_{j, k})^{P^l_{NS}(N - jk)}\prod_{1 \leq l \leq 2N}(d_{l})^{\bar{P}_{NS}(N - l/2)}.
\ee
The $f, g$, and $h$ curves are given by\footnote{We correct a small typo on p. 15, Eq. (7.10) of \cite{gepner2001unitary}. The second line of the expression for $g_{j,k}^{\text{NS}}$ should read $\frac{1}{18}\left((3+c)(1-j)+9\frac{1-k}2\right)\left((3+c)(1+j)+9\frac{1+k}2\right)$.} 
\begin{align}
f_{m, n} &= \frac{1}{192}\left(4+24x-(5m-3n)^2\right)\\
g_{j, k} &= \frac18 \left( 49 - 120h + 24x - (10j+3k)^2 \right) \\
d_{l} &= \frac{9}{64} - \frac{3h}{4} + h^2 - \frac{53 l^2}{288} + \frac{5 h l^2}{12} + \frac{25 l^4}{576} - \frac{l^2 x}6.
\end{align}

The generating functions $P_{NS}$ describe the number of states in the free-field theory. The free field representation is comprised of two bosons and their superpartners, so we have
\begin{align} 
\sum_n P_{NS}(n) q^n &= \prod_{k=1}^{\infty} \left(\frac{1 + q^{k-1/2}}{1-q^k}\right)^2 = \mc P(\tau) \nn \\ 
\sum_n \bar{P}^l_{NS}(n) q^n &= \frac{1}{(1 + q^{l/2})}\prod_{k=1}^{\infty} \left(\frac{1 + q^{k-1/2}}{1-q^k}\right)^2 = \bar{\mc{P}}^l(\tau). \label{eq:partition}
\end{align}

$\bar{\mc{P}}^{l}(\tau)$ comes when we have a fermionic level-$l$ operator that annihilates the highest-weight state: $O_{-l} \ket{h} = 0$. This is probably most familiar from the $\mathcal{N}=2$ algebra, which is discussed in more detail in Appendix B of \cite{kiritsis1988character}. There exists a null vector at level-1/2 which is annihilated by $G_{-1/2}$. Therefore, at each level we must remove basis elements containing $G_{-1/2}$ and divide out the corresponding factor, $(1 + x^{1/2})$, from the partition function. 

Something similar happens for the Spin(7) algebra. For example, $\left(\frac{5}{2} G_{-1/2} + \thalf M_{-1/2}\right)\ket{0,0}$ is a singular highest weight state annihilated by $\thalf G_{-1/2} - M_{-1/2}$. Similarly, $\left( \thalf G_{-1/2} - M_{-1/2} \right) \ket{\thalf, \thalf}$ is a singular highest weight state annihilated by $\frac{7}{2} G_{-1/2} - M_{-1/2}$. Both of these singular states produce Verma modules generated by $\bar{\mc{P}}^{1}(\tau) = \frac{\mc{P}(\tau)}{1+x^{\frac{1}{2}}}.$

\begin{figure}
\begin{center}
\includegraphics[width=0.75 \textwidth]{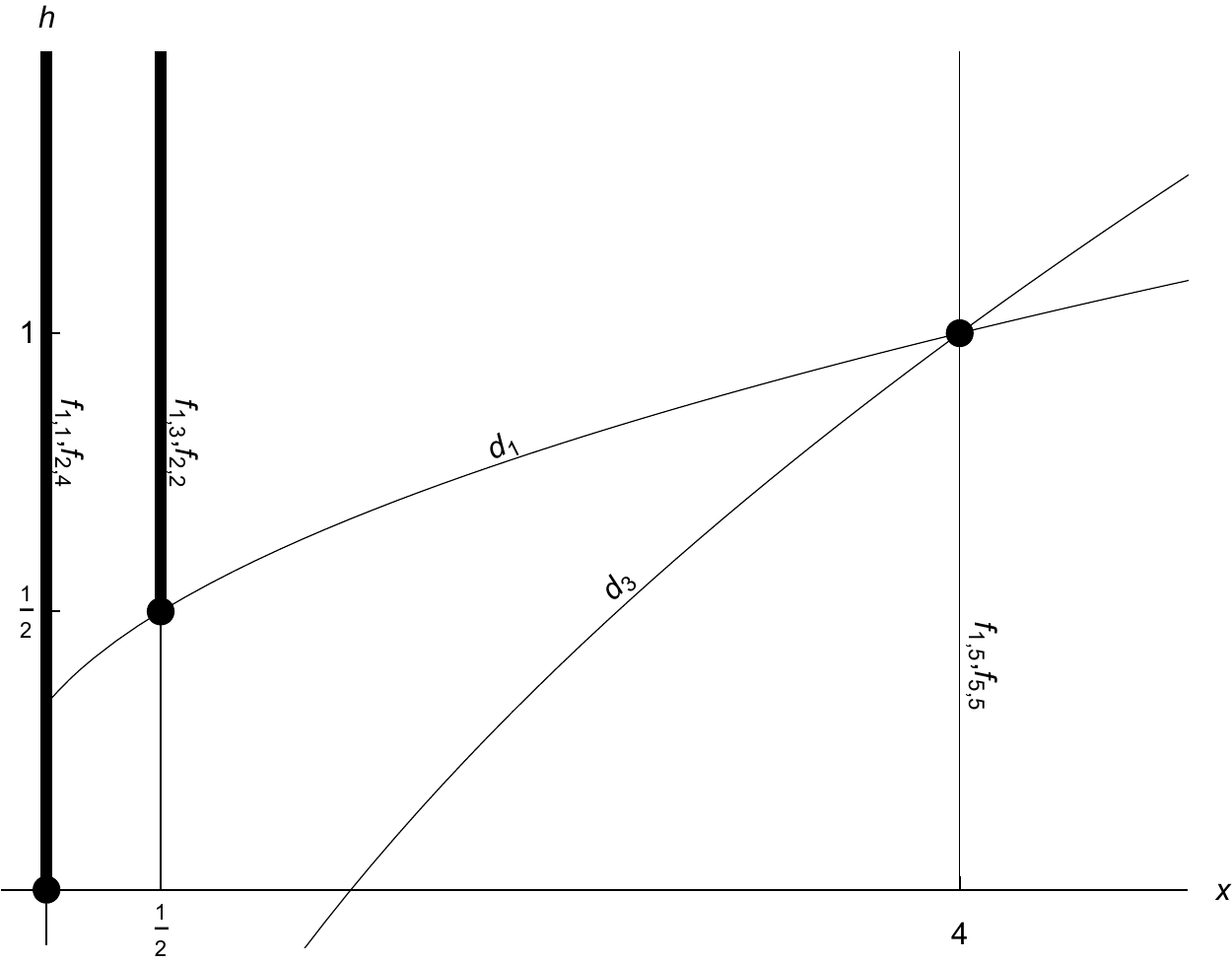}
\end{center}
\caption{The vanishing curves of the $f$, $d$, and $g$ curves in the Kac determinant in terms of the eigenvalues of the highest weight state, taken from \cite{gepner2001unitary}.  The black dots indicate the location of massless unitary modules and the two vertical black lines are continua of massive unitary modules.}
\label{fig:vanishing}
\end{figure}

In order to find the unitary representations, we need the vanishing curves of the $f, g,$ and $d$ curves.   The first few curves are plotted in Figure \ref{fig:vanishing} in terms of the $h$ and $x$, the $L_0$ and $X_0$ eigenvalues of the highest weight space respectively.  Gepner and Noyvert specified which of the modules are unitary and the results are also in the figure.

Gepner and Noyvert write down two continuous series of unitary highest weight representations that we call massive representations and three discrete representations that we call massless representations~\cite{gepner2001unitary} . The massive representations are labeled by highest weight states $\ket{h, 0}$  with $h>0$ and $\ket{h, \thalf}$ with $h>\thalf$. The massless characters are labeled by highest weight states $\ket{0,0}$ (the vacuum), $\ket{\thalf, \thalf}$, and $\ket{1,4}$.
 
\subsection{The Content of our Characters}
In this section we construct (conjectural) characters for the massive and massless unitary highest weight representations of the Spin(7) algebra in the NS sector. 

\subsubsection{Massive Characters}
In the case of the massive characters, it is sufficient to look for solutions of the Kac determinant to find the location of singular vectors, then to treat these singular vectors as new highest weight states and look for new singular vectors, and so on.   The character of the unitary representation is then found by subtracting out all singular vectors via inclusion/exclusion\footnote{We do not expect exotica like subsingular vectors to appear in the massive characters, which have, for example only a single Kac determinant vanishing curve, much like the conventional Virasoro case. This expectation is borne out by explicit numerical checks and the complementary computations of \cite{eguchi2003supercoset}. For a more precise explanation of the unusual features of $\mathcal{W}$-algebra representation theory see, e.g., \cite{bouwknegt1996w3, de1996non}.}.

As an example, consider a Verma module generated by the highest weight state $| h \rangle$. Suppose this state has two singular descendants, $|n_1 \rangle, |n_2 \rangle$, whose contributions we wish to subtract from the character. If we denote the modules generated by a highest weight vector $\lambda$ by $M(\lambda)$ then the expression for the character becomes $\operatorname{ch}M(h) - \operatorname{ch}M(n_1) - \operatorname{ch}M(n_2)$. However, if $|n_1 \rangle$ and $|n_2 \rangle$ share a singular descendant, $|n_3 \rangle$, then in subtracting $M(n_{1, 2})$ we have doubly subtracted $M(n_3)$ and therefore must add a term to our character to compensate: $$\chi = \operatorname{ch}M(h) - \operatorname{ch}M(n_1) - \operatorname{ch}M(n_2) + \operatorname{ch}M(n_3).$$ 

Let's first compute the character for $\ket{h,0}$ for some $h > 0$. The other massive character will be treated identically; all that differs is finding the particular quantum numbers.

The first thing to note is that the only vanishing curve ``intersecting" the two massive states is the $f_{m, n}$ curve in Figure~\ref{fig:vanishing}. While it is possible in general for the descendant singular vectors to admit descendants of their own that satisfy the $g_{j, k}$ or $d_{l}$ equations (and this will happen in the massless case), it does not happen here. 

The characters will in general have the form
\be 
\chi_{h,0} = q^{h-\half}\mc P(\tau) (1- f_{\text{singular}}),
\ee
where $\mc P(\tau)$ is defined in (\ref{eq:partition}) and $f_{\text{singular}}$ is the contribution of all the singular vectors.

We plug $x=0, h>0$ into $f_{m, n}$ and search for solutions that satisfy $m, n \in \mathbb{Z}, m + n \in 2 \mathbb{Z}$.  We find two sets of solutions:
\begin{align}
n_{1,k} &= (3k + 2), \ m_{1,k} = (5 k + 4), \ k= 0, 1, \ldots \nn \\
n_{2,k} &= (3k + 1), \ m_{2,k}= (5 k + 1),  \ k= 0, 1, \ldots
\end{align}
Because these new singular vectors are all by themselves highest weight, they generate their own Verma module from which we can find new singular vectors, by finding solutions to $f_{m,n}=0$ and plugging in $h + \frac{n_{i,k} m_{i,k}}{2}$, $i = 1, 2$. For example, using as our highest weight vector the singular vector at level $\half$ from the first set of solutions ($i=2$ and $k=0$) we find:
\begin{align}
n_{1,j}&= (3j + 2), \ m_{1,j} = (5 j + 6), \ j= 0, 1, \ldots \nn \\
n_{2,j} &= (3j + 4), \ m_{2,j}= (5 j + 4),  \ j= 0, 1, \ldots
\end{align}

\begin{figure}[h]
\begin{center}
\includegraphics[width=0.48\textwidth]{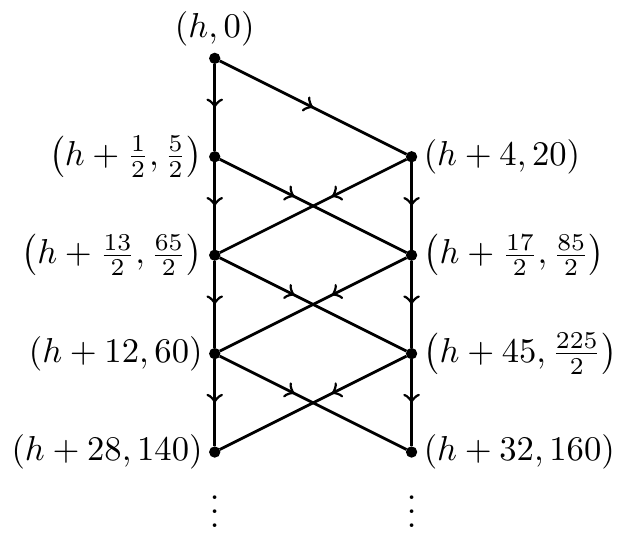}
\end{center}
\caption{The embedding diagram that determines the character $\chi_{h,0}$.
}
\label{fig:0hmass}
\end{figure}

After iterating this procedure, we exhaust the singular vectors and organize the result in the embedding diagram in Figure~\ref{fig:0hmass}.  Nodes in this diagram are singular vectors with the indicated set of quantum numbers under $L_0$ and $X_0$.  Arrows indicate descendants; there is an arrow from some singular vector $v$ to $w$ if $w$ is in the submodule generated by $v$.

\begin{figure}[h]
\begin{center}
\includegraphics[width=0.48\textwidth]{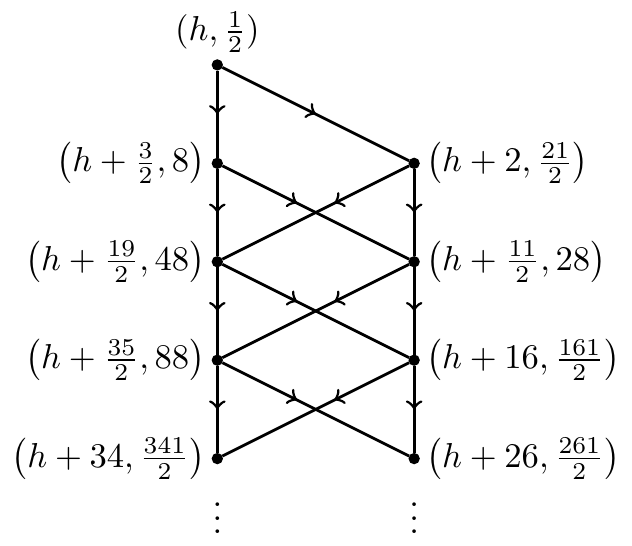}
\end{center}
\caption{The embedding diagram that determines the character $\chi_{h,\frac{1}{2}}$.
}
\label{fig:halfhmass}
\end{figure}

The embedding structure is necessary to determine how to subtract all singular vectors, to prevent overcounting. The answer for $\chi_{h,0}$ is:
\be \label{eq:mass1}
\chi_{h,0}= q^{h-\half} \mc P(\tau) \left ( 1- \sum_{k=0}^\infty q^{(3k+1)(5k +1)\over 2}- \sum_{k=0}^\infty q^{(3k+2)(5k +4)\over 2}+ \sum_{k=0}^\infty q^{(3k+2)(5k +6)+1\over 2}+\sum_{k=0}^\infty q^{(3k+4)(5k +4)+1\over 2}\right ).
\ee
After repeating the same procedure for the other massive tower, we find:
\be\label{eq:mass2}
\chi_{h,\frac{1}{2}}= q^{h-\half} \mc P(\tau) \left ( 1- \sum_{k=0}^\infty q^{(3k+1)(5k +3)\over 2}- \sum_{k=0}^\infty q^{(3k+2)(5k +2)\over 2}+ \sum_{k=0}^\infty q^{(3k+1)(5k +7)+4\over 2}+\sum_{k=0}^\infty q^{(3k+5)(5k +3)+4\over 2}\right ).
\ee
whose embedding diagram is shown in Figure \ref{fig:halfhmass}.

These answers agree with those derived in \cite{eguchi2003supercoset}, who employed a coset construction of the algebra.
 
\subsubsection{Massless Characters}

We now turn to the three discrete massless unitary highest weight representations.  For these computations, the Kac determinant proves to be insufficient to obtain the correct character formulae. To assist us, we employ numeric methods that compute the characters to finite order by explicitly constructing the algebra. The Mathematica code and the details of the algorithm are reported in~\cite{whalen2014algorithm}.

The Kac determinant can fail to provide complete information about the characters in the following four ways \cite{bowcock1997representation, bowcock1998characters}:
\begin{enumerate}
\item The Kac determinant may propose states that evaluate to zero.
\item The Kac determinant may fail to identify the complete embedding structure among Verma modules by failing to find arrows in the embedding diagram.
\item The Kac determinant will not provide information about multiplicity of states with identical eigenvalues.
\item If there exist null states that are descendants of a unitary highest weight state that are neither highest weight themselves (singular) nor descendants of singular vectors, the Kac determinant will fail to find them. These states may become singular in the quotient module constructed by modding out the original highest weight module by all singular vectors, and are sometimes called subsingular vectors\footnote{This is closely related to the mathematical notion of a primitive vector, a vector $v$ along with a submodule $U$ of the entire module $V$ such that $v\not\in U$, but $v+U$ is singular in $V/U$.  A subsingular vector is a primitive vector where $U$ is the space generated by singular vectors in $V$.}.
\end{enumerate}  
The fourth item is frequently assumed not to occur in computations of characters (and in many of the most physically interesting algebras, it does not). Unhappily, these states seem to appear in two of our three massless characters and are generally quite complicated. We present an explicit subsingular vector in  Appendix \ref{app:subsingular}. Though we believe we have found all such vectors, it would be desirable to have a proof of this and, more generally, a systematic analytical way for finding all subsingular vectors, including those at very high levels inaccessible to numerics\footnote{We are grateful to Daniel Bump and Valentin Buciumas for preliminary discussions on this point.}.

The first three subtleties may be dealt with systematically by explicitly constructing the offending states. This has been done to great effect in, for example, \cite{bowcock1997representation, bowcock1998characters}. We proceed numerically in a similar spirit, supplementing this approach with more standard computations from the Kac determinant as described above. In the course of these computations, we  find the following BPS-like relations are satisfied by the characters:
\begin{align} \label{eq:bps}
\tilde{\chi}_{0,0} + \tilde{\chi}_{\half,\half} &=q^{-n} \chi_{n,0} \nn \\
\tilde{\chi}_{\frac{1}{2},\half} + \tilde{\chi}_{1,4} &=q^{-n}  \chi_{n+\half,\half}
\end{align}
where $\tilde{\chi}_{h,x}$ denotes the massless characters with internal Ising weight $\frac{x}8$. We will derive these equations in \S\ref{sec:ramond}, but for now let us motivate them heuristically. Very roughly, one can see this at the level of the Kac determinant and embedding diagrams by noticing the following. The massive towers only possess singular vectors generating full Verma modules via $\mc P(\tau)$, coming from $f_{m, n}$ solutions while the massless towers possess additional $d_{l}$ solutions which generate ``truncated" modules (i.e. modules with partition function $\bar{\mc{P}}^{l}(\tau)$). However, two $d_{l}$ Verma modules can sum up to a contribution equivalent to a single $f_{m, n}$ Verma module. One can track this elaborate series of splittings in the embedding diagrams of all the characters and convince oneself that, at least up to a certain order in $q$, these relations are satisfied. The analogue of this in the $\mathcal{N}=2$ case is that the short Verma module comes from BPS states satisfying $G_{-1/2}|state \rangle = 0$ which then have truncated modules as described earlier. However, as is commonly known, one long multiplet can split into two short multiplets at the $h \rightarrow q/2$ threshold. 

\begin{figure}
\begin{center}
\includegraphics[width=0.7\textwidth]{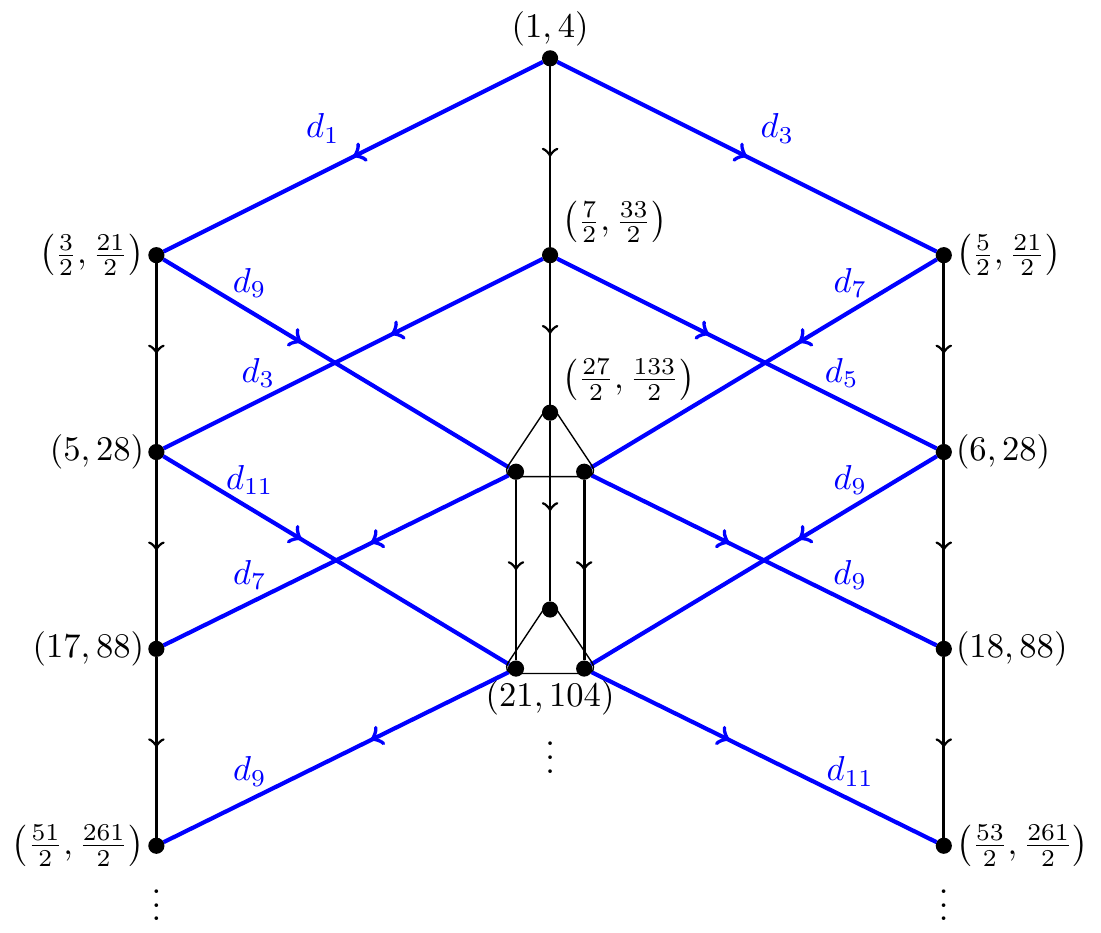}
\end{center}
\caption{The embedding diagram for the massless character $\tilde \chi_{1,4}$.  Blue arrows 
represent d-curve descendants.  The triangles are written in the notation of D\"orrzapf\cite{dorrzapf1998embedding}: each triangle and the three dots represent a two-dimensional space of singular vectors with the same quantum numbers.  The three dots each represent a singular vector that generates the one-dimensional intersection of the singular space with the descendants of the source of the arrow.  These dots are pairwise linearly independent.
}

\label{fig:1h}
\end{figure}

\begin{sidewaysfigure}
\begin{center}
\includegraphics[width=0.9\textwidth]{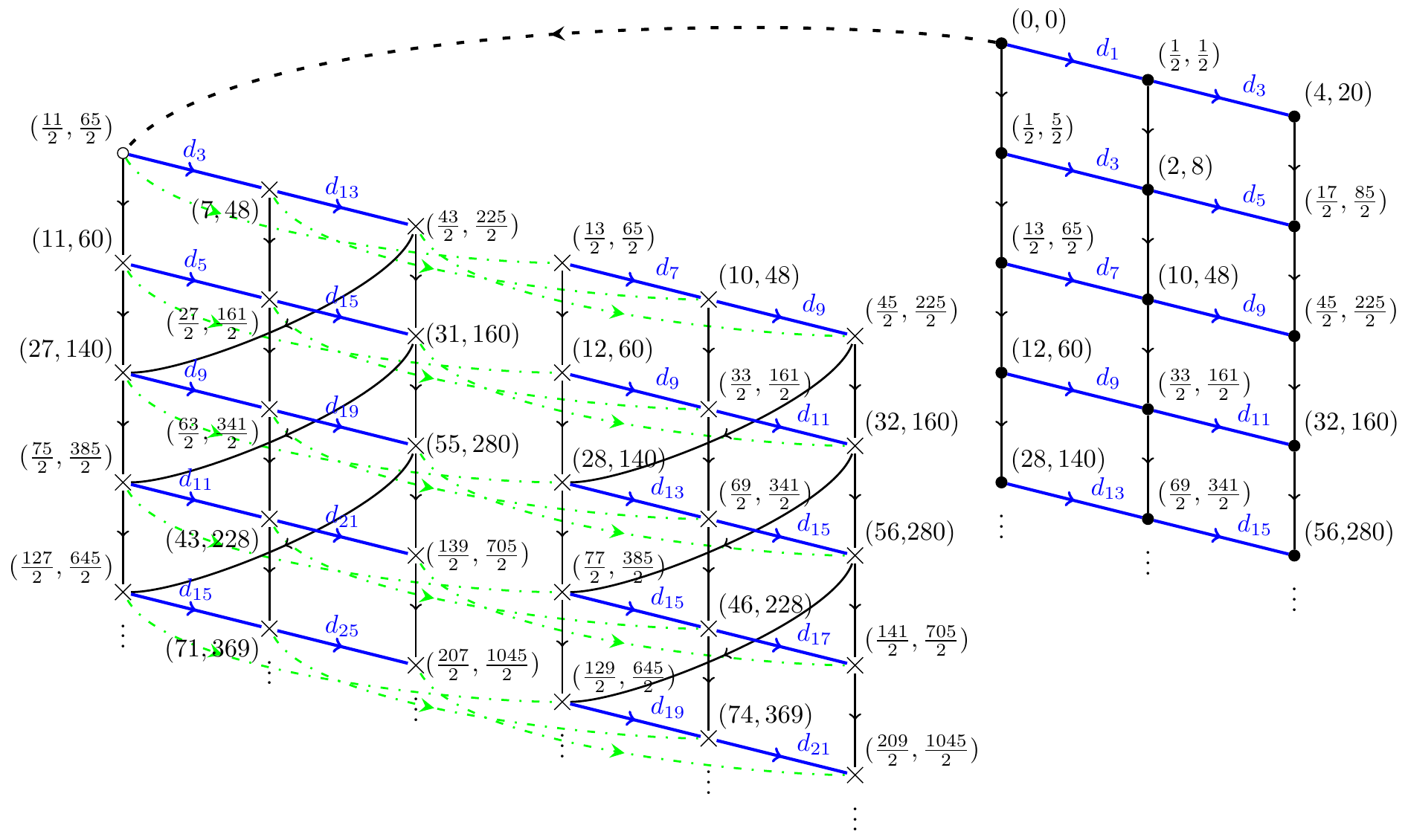}
\end{center}
\caption{The embedding diagram which determines the massless character $\tilde \chi_{0,0}$. The green lines represent solutions to $g_{1,1}=0$ and $g_{1,3}=0$ in the Kac determinant.
}
\label{fig:0h}
\end{sidewaysfigure}

\begin{sidewaysfigure}
\begin{center}
\includegraphics[width=0.7\textwidth]{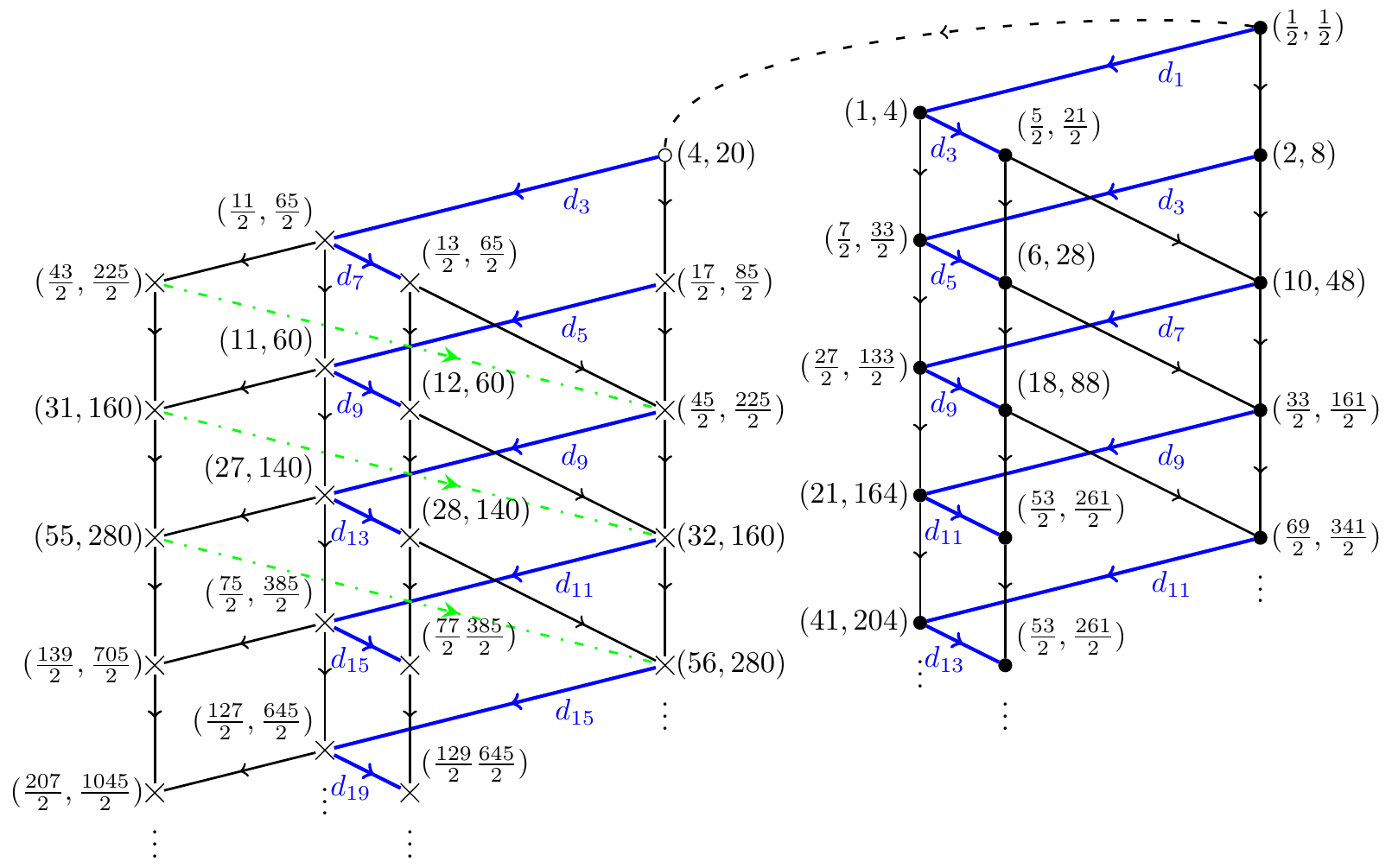}
\end{center}
\caption{The embedding diagram which determines the massless character $\tilde \chi_{\half,\half}$.  The green arrows are $g$-curves.  The rightward arrows in the subsingular diagram are actually $g$-curves in the Kac-determinant, but since they follow a $d$-curve, they induce $d$-type descendants in the target singular vector as though they were the $d$-curve indicated in the diagram.
}
\label{fig:halfh}
\end{sidewaysfigure}

Below, we present the embedding diagrams (to finite level) obtained by explicitly constructing null states and, in the case of all singular vectors, determining their parentage (i.e. from which highest weight vector they descended).  These appear in Figure \ref{fig:1h}, \ref{fig:0h}, and \ref{fig:halfh}.

Figure \ref{fig:0h} and \ref{fig:halfh} use additional formalism from the massive case.  Given a module $V$, the diagram is separated into two subdiagrams.  The subdiagram on the right is the space $U$ of singular vectors as in the massless case.  The circle on the left represents a subsingular vector in $V$, that is, a singular vector in $V/U$.  The tower on the left is the diagram for the Verma module $W$ generated by a highest weight vector with the same quantum numbers as the subsingular vector.  The  $\times$s on the left are the singular vectors in $W$ that are in the kernel of the induced homomorphism $W\to U/V$.

We conjecture closed form expressions for the character formulae by extrapolating the structure of the diagrams to higher level, imposing satisfaction of (\ref{eq:bps}) and, when possible, finding singular vector locations of the Kac determinant in closed form. Our confidence in our character formulae is bolstered by the agreement of our massive characters with those in \cite{eguchi2003supercoset}.

We conjecture the massless characters $\tilde{\chi}_{h,x}$ to be:
\begin{align}
\tilde{\chi}_{0,0} &= q^{-\half} \mc P(\tau) \Bigg(1 - \sum _{k=0}^\infty \bigg(q^{\frac{15}2k^2+4k+\half}+\frac{q^{\frac{15}2k^2 +2k+\half}}{1+q^{\frac{6 k+1}2}}-\frac{q^{\frac{15}2k^2+7k+2}}{1+q^{\frac{6 k+3}2}} \nn \\
&\phantom{aaaaaaaaaaaaaaaaaaaaa} -q^{\frac{15}2k^2+14k+\frac{13}2}+\frac{q^{\frac{15}2k^2+14k+\frac{11}2}}{1+q^{\frac{6k+3}2}}-\frac{q^{\frac{15}2k^2+19k+11}}{1+q^{\frac{6 k+5}2}}\bigg)\Bigg) \nonumber \\
\tilde{\chi}_{\half,\half} &= \mc P(\tau) \Bigg( 1 - \sum _{k=0}^\infty \bigg(q^{\frac{15}2k^2+7k+\frac32}-\frac{q^{\frac{15}2k^2+16k+8}}{1+q^{\frac{6 k+5}2}}+ \frac{q^{\frac{15}2 k^2 + 5k + \half}}{1+q^{\frac{6 k+1}2}} \nn \\
&\phantom{aaaaaaaaaaaaaaaaaaaaa}-q^{\frac{15}2 k^2 + 17k + \frac{19}{2}}+\frac{q^{\frac{15}2 k^2 + 11k + \frac72}}{1+q^{\frac{6k+3}2}}-\frac{q^{\frac{15}2 k^2 + 10k + 3}}{1+q^{\frac{6 k+3}2}}\bigg)\Bigg ) \nonumber \\
\tilde{\chi}_{1,4} &= q^\half \mc P(\tau)\Bigg (1-\sum_{k=0}^\infty \bigg (q^{\frac{15}2k^2+10k+\frac52}-q^{\frac{15}2k^2+20k+\frac{25}2}+\frac{q^{\frac{15}2 k^2 + 8 k + \half}}{1 + q^{6 k + 1\over2}} \nn \\
&\phantom{aaaaaaaaaaaaaaaaaaaaa}- \frac{q^{{15\over2} k^2 + 13 k + 4}}{1 + q^{6 k + 3\over2}} + \frac{q^{{15\over 2} k^2 + 8 k + \frac32}}{1 + q^{6 k + 3\over 2}} - \frac{ q^{{15\over 2} k^2 + 13 k + 5}}{1 + q^{6 k + 5\over 2}}\bigg )\Bigg ). 
\end{align}

\subsection{What Kind of Mockery is This?}  
We would like to understand the modular properties of our conjectural NS-sector characters, and we will do so by relating them to standard modular and mock modular forms of a single variable. In \S\ref{sec:massless} we will discuss the mock modular properties of the massless characters, and in \S\ref{sec:massive} we will show how the massive characters transform as a two-component vector-valued modular form under $\Gamma_{\theta}$\footnote{Some of the following (mock) modular identities are also discussed in \cite{M5spin7}, in which the characters are labeled by their Ising weights $a$, rather than $x=8a$. We will label quantities by $x$ throughout.}.
\subsubsection{Massless characters}\label{sec:massless}
First we define some useful functions,
\be
\theta_{m,r}(\tau,z)  = \sum_{k =r\!\pmod{2m}} q^{\frac{k^2}{4m}}y^{2mk}
\ee
and
\be\label{eq:AL}
f_u^{(m)}(\tau,z)= \sum_{k\in \mathbb Z} \frac{q^{mk^2} y^{2mk}}{1- yq^k e^{-2\pi iu}}.
\ee
Note that $f_u^{(m)}$ has the so-called elliptic transformation property for $2m\in\ZZ$: 
\[
f_u^{(m)}(\tau,z)  = f_u^{(m)}(\tau,z+1) = q^m y^{2m} f_u^{(m)}(\tau,z+\t) ,
\]
and $\theta_{m,r}(\t,z)=\theta_{m,-r}(\t,-z)$. We will also make use of a single variable version of this function
\be
\theta_{m,r}(\tau)  = \theta_{m,r}(\tau,0),
\ee
which is what we mean whenever $z$ is suppressed, satisfying $\theta_{m,r}(\t)= \theta_{m,-r}(\t) = \theta_{m,r+2m}(\t)$, and
\be
\til \th_{m,r} (\t)= \th_{m,r} (\t) + \th_{m,r-m} (\t)  
\ee
satisfying $\til \th_{m,r}=\til \th_{m,-r}=\til \th_{m,r+m}$.
In \cite{Zwegers} it was shown that one can define a (non-holomorphic) completion of $f^{(m)}_u(\tau,z)$
\be\label{eq:completion}
\hat f^{(m)}_u(\tau,\overline \tau,z)=f^{(m)}_u(\tau,z) - {1\over 2}\sum_{r\bmod 2m} R_{m,r}(\tau,u) \theta_{m,r}(\tau,z)
\ee
which transforms as a Jacobi form of weight 1 and index $m$. Here we have defined
\be
R_{m,r}(\tau,u)= \sum_{k=r\bmod 2m}\left ( \sgn\left (k + {1\over 2}\right ) - E\left( \left(k + 2m {\Im u\over \Im \tau}\right)\sqrt{\Im \tau\over m}\right )\right)q^{-{k^2\over 4m}}e^{-2\pi i k u}
\ee
where
\be
E(z) = \sgn(z)\left (1- \int_{z^2}^\infty dt~ t^{-{1\over 2}}e^{-\pi t}\right ).
\ee

We can rewrite the massless characters in terms of specializations of the function (\ref{eq:AL}) to particular values of $y$ and $u$, as
\begin{multline}
\tilde{\chi}_{0,0}=\frac{q^{-{3\over 8}}\eta(\tau)^2}{\eta\left ({\tau\over 2}\right)^2 \eta(2\tau)^2}\Bigg (f^{(5)}_{-{7\tau\over 10} + {1\over 2}}\left(6\tau,-{\tau\over 5}\right)+q^{13\over 2}f^{(5)}_{-{7\tau\over 10} + {1\over 2}}\left(6\tau,{14\tau\over 5}\right)-q^{1\over 2}f^{(5)}_{-{7\tau\over 10} + {1\over 2}}\left(6\tau,{4\tau\over 5}\right) \\ -q^4 f^{(5)}_{-{7\tau\over 10} + {1\over 2}}\left(6\tau,-{11\tau\over 5}\right)\Bigg),
\end{multline}
\begin{multline}
\tilde{\chi}_{\half,\half}=  \frac{q^{1\over 8}\eta(\tau)^2}{\eta\left ({\tau\over 2}\right)^2 \eta(2\tau)^2}\Bigg( f^{(5)}_{-{\tau\over 10} + {1\over 2}}\left(6\tau,{2\tau\over 5}\right)+q^{11\over 2}f^{(5)}_{-{\tau\over 10} + {1\over 2}}\left(6\tau,-{13\tau\over 5}\right)-f^{(5)}_{{7\tau\over 10} + {1\over 2}}\left(6\tau,-{4\tau\over 5}\right) \\ -q^{7\over 2}f^{(5)}_{{7\tau\over 10} + {1\over 2}}\left(6\tau,{11\tau\over 5}\right)\Bigg ),
\end{multline}
\be
\tilde{\chi}_{1,4}=  \frac{\eta(\tau)^2}{\eta\left ({\tau\over 2}\right)^2 \eta(2\tau)^2}\Bigg( q^{5\over 8}\left (f^{(5)}_{{\tau\over 2} + {1\over 2}}(6\tau,\tau)-f^{(5)}_{{\tau\over 2} + {1\over 2}}(6\tau,-\tau)\right )+q^{25\over 8}\left (f^{(5)}_{{\tau\over 2} + {1\over 2}}(6\tau,-2\tau)-f^{(5)}_{{\tau\over 2} + {1\over 2}}(6\tau,2\tau)\right )\Bigg ).
\ee
Here we have used that
\be\nonumber
\mc P(\t)=\frac{\eta(\tau)^2}{\eta(\tfrac{\t}{2})^2 \eta(2\tau)^2}.
\ee
Note that $\mc P(\t)$ transforms as a weight $-1$ modular form under the subgroup $\Gamma_\theta$.
Each of these characters is composed of holomorphic two-component vector-valued mock modular forms which can be completed into  non-holomorphic (two-component, vector-valued) modular forms via equation (\ref{eq:completion}) with some specialization of $u$ and $y$. For example, for the case of $\tilde \chi_{1,4}$, defining
\be
\overline \mu^{NS}=q^{5\over 8}\left (f^{(5)}_{{\tau\over 2} + {1\over 2}}(6\tau,\tau)-f^{(5)}_{{\tau\over 2} + {1\over 2}}(6\tau,-\tau)\right )+q^{25\over 8}\left (f^{(5)}_{{\tau\over 2} + {1\over 2}}(6\tau,-2\tau)-f^{(5)}_{{\tau\over 2} + {1\over 2}}(6\tau,2\tau)\right ),
\ee
we see that we can define a completion, $\hat{\overline \mu}^{NS}(\t,\overline\t)$, 
\be
\hat{\overline \mu}^{NS}(\tau,\overline \tau)=\overline\mu^{NS}(\tau)-{1\over 2}\frac{1}{\sqrt{60 i}}\int_{-\overline \tau}^{i\infty} d\tau'~(\tau' + \tau)^{-{1\over 2}}\ubar{\theta}_{NS}(\t) \cdot \ubar{S}(\tau')
\ee
which transforms as a weight $1$ modular form under $\Gamma_{\theta}$,\footnote{We thank Miranda Cheng for finding an error in an earlier version of this section.}
where we have defined
\[ \nonumber\ubar{\theta}_{NS}(\t)=\left( \begin{array}{c}
\Theta^{NS}_{1/2}(\t) \\
\Theta^{NS}_{0}(\t)  \end{array} \right),\]
\begin{eqnarray}\label{eq:thetasNS}
\Theta^{NS}_{0}(\t) &\equiv \til\th_{30,2}(\t)-\til\th_{30,8}(\t) = \sum_{k \in \bb Z}\epsilon^{NS}_{0}(k)q^{k^2 \over 120} \\
\Theta^{NS}_{1/2}(\t)&\equiv \til\th_{30,4}(\t)-\til\th_{30,14}(\t)= \sum_{k \in \bb Z}\epsilon^{NS}_{1/2}(k)q^{k^2 \over 120},
\end{eqnarray}
with
\begin{displaymath}
   \epsilon^{NS}_{0}(x) = \left\{
     \begin{array}{lr}
       1 &  k= 2, 28 \ (\textrm{mod} \ 60)\\
      -1 &  k= -8, -22 \ (\textrm{mod} \ 60) \\
       0 & \textrm{otherwise}
     \end{array}
   \right.
\end{displaymath}
\begin{displaymath}
   \epsilon^{NS}_{1/2}(k) = \left\{
     \begin{array}{lr}
       1 &  k= 4, 26 \ (\textrm{mod} \ 60)\\
      -1 &  k= -14, -16 \ (\textrm{mod} \ 60) \\
       0 & \textrm{otherwise}
     \end{array},
   \right.
\end{displaymath}
\[\nonumber \ubar S(\t)=\left( \begin{array}{c}
S_1 \\
S_7  \end{array} \right),\]
and finally,
\be\label{eq:shadow}
S_{\alpha}(\tau)= \sum_{k \in \bb Z}k \epsilon^{R}_{\alpha}(k) q^{k^2 \over 120}, \  \alpha = 1, 7
\ee
satisfying 
\begin{displaymath} \label{epsilonR1}
   \epsilon^{R}_{1}(x) = \left\{
     \begin{array}{lr}
       1 &  k= 1, 29 \ (\textrm{mod} \ 60)\\
      -1 &  k= -11, -19 \ (\textrm{mod} \ 60) \\
       0 & \textrm{otherwise}
     \end{array}
   \right.
\end{displaymath}
\begin{displaymath} \label{epsilonR7}
   \epsilon^{R}_{7}(k) = \left\{
     \begin{array}{lr}
       1 &  k= -7, -23 \ (\textrm{mod} \ 60)\\
      -1 &  k= 17, 13 \ (\textrm{mod} \ 60) \\
       0 & \textrm{otherwise}
     \end{array}.
   \right.
\end{displaymath}
Though seemingly unwieldy, these definitions prove useful when deriving transformations under the modular group. Including the factor of $\mc P(\t)$, the character $\tilde \chi_{1,4}=\mc P(\t)\overline \mu^{NS}$ as a whole transforms as a weight $0$ mock modular form under~$\Gamma_\theta$.

\subsubsection{Massive characters}\label{sec:massive}
Now we'd like to discuss the modular properties of the massive characters.
After some mathematical manipulation, the characters for the massive states in the NS sector given in equations (\ref{eq:mass1}) and (\ref{eq:mass2}) can then be written as
\be
\chi_{h,0}=q^{h-{49\over 120}} \frac{\eta(\tau)^2}{\eta(\tfrac{\t}{2})^2 \eta(2\tau)^2} \Big(\til\th_{30,2}(\t)-\til\th_{30,8}(\t)\Big) \equiv q^{h-{49\over 120}} \mc P(\t) \Theta^{NS}_{0}
\ee
and
\be
\chi_{h,{1\over 2}}=q^{h-{61\over 120}} \frac{\eta(\tau)^2}{\eta(\tfrac{\t}{2})^2 \eta(2\tau)^2} \Big(\til\th_{30,4}(\t)-\til\th_{30,14}(\t)\Big) \equiv q^{h-{61\over 120}}\mc P(\t) \Theta^{NS}_{1/2}.
\ee
We will show that the theta functions appearing in these characters transform as a two-component vector under $\Gamma_{\theta}$. 

First consider how $\theta_{m,r}(\tau)$ for a fixed $m$ and $r$ transforms under $T:\tau \mapsto \t + 1$ and $S: \t\mapsto -1/\t$. We have
\be\label{eq:Ttrans}
\theta_{m,r}(\tau+1)= e\left ( {r^2\over 4m}\right ) \theta_{m,r}(\t)
\ee
for the $T$ transformation where we use the shorthand $e(x)=e^{2\pi i x}$. For the S transformation, we have
\bea\nonumber
\theta_{m,r}\left (-{1\over \tau}\right ) &=& \sum_{k\in \mathbb Z}e^{-{2\pi i\over \t}\left (mk^2 + rk+ {r^2\over4m}\right)}\\\nonumber
&=&e\left (-{r^2 \over 4m\t}\right ) \sum_{k\in \mathbb Z}\int_{-\infty}^\infty dx ~e^{2\pi i x k}e^{-{2\pi i\over \t}(mx^2 + rx)}\\ \label{eq:Stransv1}
&=&\sqrt{-i\t\over 2m} \sum_{k\in \mathbb Z}e^{2\pi i \t k^2\over 4m}e\left(-{rk\over 2m}\right)
\eea
where in the second line we have used the Poisson transformation formula, and in the third we used
\be
\int_{-\infty}^\infty dx~ e^{2\pi i xy} e^{-tx^2}= {e^{-\pi y^2/t}\over \sqrt{t}}.
\ee
Now let's try to write equation (\ref{eq:Stransv1}) in terms of theta functions of the same $m$ but different $r$. Note that
\be
\sum_{k\in \mathbb Z}e^{2\pi i \t k^2\over 4m}e\left(-{rk\over 2m}\right)= \sum_{r'=-m+1}^{m}\sum_{k \in \mathbb Z}e^{2\pi i\t \left(mk^2 + r'k +{ r'^2\over4m}\right)}e\left (-{rr'\over2m}\right),
\ee
so finally\footnote{This S-transformation formula is often repackaged in the literature by defining $\mc S^{(\theta)}_{r r'} \equiv \frac{1}{\sqrt{2m}} e^{\frac{i \pi r r'}{m}}$ so that $\theta_{m,r}\left (-{1\over \tau}\right )=\sqrt{-i\t} \mc S^{(\theta)} \theta_{m}(\t, z)$. Similarly, one often defines $\mc T^{\theta}_{r r'} \equiv e^{\pi i r^2 \over 2m} \delta_{r, r'}$ so $\theta_{m}(\t + 1, z)= \mc T^{(\theta)} \theta_{m}(\t, z)$.}
\be\label{eq:Strans}
\theta_{m,r}\left (-{1\over \tau}\right )=\sqrt{-i\t\over 2m}\sum_{r'=-m+1}^{m}e\left (-{rr'\over2m}\right)\theta_{m,r'}(\t).
\ee
Using formulas (\ref{eq:Ttrans}) and (\ref{eq:Strans}), the vector $\ubar{\theta}_{NS}(\t)$ defined in the previous section transforms in the following way under the generators of $\Gamma_{\theta}$, $T^2$ and $S$:
\be
\ubar{\theta}_{NS}(\t+2)= \rho(T^2)\cdot \ubar{\theta}_{NS}(\t), ~~\ubar{\theta}_{NS}\left(-{1\over\t}\right)= \sqrt{\tau} \rho(S)\cdot \ubar{\theta}_{NS}(\t)
\ee
where

\[\rho(T^2)= \left( \begin{array}{cc}
e\left ({32\over 120}\right )& 0  \\
0 & e\left ({8\over 120}\right )  \end{array} \right)\]
and
\[\rho(S)= e \left(- \frac{1}{8} \right) \left(
\begin{array}{cc}
 -\sqrt{\frac{2}{5+\sqrt{5}}} & \sqrt{\frac{2}{5-\sqrt{5}}} \\
 \sqrt{\frac{2}{5-\sqrt{5}}} & \sqrt{\frac{2}{5+\sqrt{5}}} \\
\end{array}
\right).
\]

\subsection{Characters in the Ramond sector}\label{sec:ramond}
In the Ramond sector, the massive states have two components (see \cite{gepner2001unitary}) and are labeled by 8 times the Ising dimensions $(x_1,x_2)= (8 a_1, 8 a_2)$, and total dimension $h$. The isomorphism between NS and Ramond sector states is given explictly by in Table \ref{tbl:NSR},  where states are labeled by $|8 a,h\rangle$ and the isomorphism identifies the states in each row. Recall that this ``spectral flow" isomorphism is generated by the internal Ising sector. Specifically, the Ramond ground state with its total weight equal to its Ising weight, $\frac{1}{2}$, is isomorphic to the Ising energy operator $\left[ \epsilon \right]$ and fusion of this operator with other states generates a flow to the NS-sector. Crucially, fusion of this operator with itself gives the identity. For convenience, we reproduce the Ising fusion rules in terms of the dimension $\frac{1}{16}$ operator $\sigma$, the identity $0$, and the dimension $\half$ operator $\epsilon$:
\begin{align}
\left[ \epsilon  \right] \left[ \epsilon  \right] & = \left[ 0 \right] \nn\\
\left[ \epsilon  \right] \left[ \sigma  \right] & = \left[ \sigma \right] \nn\\
\left[ \sigma  \right] \left[ \sigma  \right] & = \left[ 0 \right] + \left[ \epsilon \right].
\end{align}
\begin{table}[htb]\begin{center}
\begin{tabular}{ccc}
NS& &R\\\hline
$\left |0,0\right\rangle$ &$ \iff $& $\left |{4},{1 \over 2}\right \rangle$\\
$\left |\frac{1}{2},{1 \over 2}\right\rangle$ &$ \iff $& $\left |{1\over 2},{1 \over 2}\right \rangle$\\
$\left |{4},1\right\rangle$ &$ \iff$ & $\left |0,{1\over 2}\right \rangle$\\
$\left |0, x\right\rangle$ &$ \iff$ & $\left |\left ({1\over 2},{4}\right ),{1\over 2}+x\right \rangle$\\
$\left |{1\over 2},{1\over 2} + x\right\rangle$ &$ \iff$ & $\left |\left (0,{1\over 2}\right ),{1\over 2}+x\right \rangle$\\
\end{tabular}\caption{The unitary irreducible highest weight representations of the Spin(7) algebra. States in the same row are isomorphic. Recall that we are labeling the states by their eigenvalues $|x_0, h \rangle = |8 a_0, h \rangle$.}\label{tbl:NSR}\end{center}
\end{table}

Following the same steps as described above for the derivation of the characters in the NS sector, we use the Ramond sector Kac determinant derived in  \cite{gepner2001unitary} to conjecture the following characters in the Ramond sector:
\be
\chi_{\left(0,{1\over 2}\right ),h + {1 \over 2}}^R=2q^{h-{1\over 120}}\frac{\eta(2\tau)^2}{\eta(\tau)^4}\left(\tilde \theta_{30,1}(\t)-\tilde \theta_{30,11}(\t)\right) \equiv 2q^{h-{1\over 120}}\frac{\eta(2\tau)^2}{\eta(\tau)^4} \Theta^{R}_{(0, 1/2)}
\ee
and
\be
\chi_{\left ({1\over 2},{4}\right ),h + {1\over 2}}^R=2q^{h-{49 \over 120}}\frac{\eta(2\tau)^2}{\eta(\tau)^4}\left(\tilde\theta_{30,7}(\t)-\tilde\theta_{30,17}(\t)\right) \equiv 2q^{h-{49\over 120}}\frac{\eta(2\tau)^2}{\eta(\tau)^4} \Theta^{R}_{(1/2, 4)}.
\ee
We can rewrite $\Theta^R(\t)$ for convenience as
\begin{eqnarray}\label{eq:thetasR}
\Theta^{R}_{(0, 1/2)}(\t) & = \sum_{k \in \bb Z}\epsilon^{R}_{1}(k)q^{k^2 \over 120} \\
\Theta^{R}_{(1/2, 4)}(\t)& = \sum_{k \in \bb Z}\epsilon^{R}_{7}(k)q^{k^2 \over 120},
\end{eqnarray}
with $\epsilon^{R}_{1, 7}(k)$ as in \ref{epsilonR1}.

Spectral flow relates $\chi_{\left(0,{1\over 2}\right ),h}^R \iff \chi^{NS}_{{1\over 2},h}$ and $\chi_{ \left({1\over 2},{4}\right ),h}^R\iff \chi^{NS}_{0,h}$. Let us pause here to derive this correspondence, as it will be a prerequisite to confirming (\ref{eq:bps}). In the massive sector, as in the massless sector, we must identify the primary state that does not change the external (non-Ising) dimension of a state with which it fuses, and which produces the identity upon fusion with itself. This is perhaps clearer when writing out the external dimensions of each component, which then sum with each Ising dimension to equal the total $h$. The unique candidate is then $| ({1\over 2},{4}),{1\over 2}+x\rangle$ which is necessarily mapped to $|0, x \rangle$ in the NS-sector. Note that the vector of external dimensions for this state is $(\frac{7}{16} + x, x)$, so the second component of the state has the properties we require. Similarly, fusion of this state with the other massive state, $|(0, {1 \over 2}), {1 \over 2} + x \rangle$ produces $|{1 \over 2}, {1 \over 2} + x \rangle$ in the NS-sector.

The important massless character will be $\left |0,{1\over 2}\right \rangle$ which we denote as  $\tilde\chi_0^R$. This is given by
\be
\tilde\chi_0^R=2\frac{\eta(2\tau)^2}{\eta(\tau)^4}\left (f^{(5)}_{{\tau\over 2} + {1\over 2}}\left(6\tau,{\tau\over 2}\right)-f^{(5)}_{{\tau\over 2} + {1\over 2}}\left(6\tau,-{\tau\over 2}\right)+q^5f^{(5)}_{{\tau\over 2} + {1\over 2}}\left(6\tau,-{5\tau\over 2}\right)-q^5f^{(5)}_{{\tau\over 2} + {1\over 2}}\left(6\tau,{5\tau\over 2}\right)
\right ).
\ee
Spectral flow relates this character to $\tilde\chi_{1,4}$ . The Ramond sector BPS relations 
\be\label{eq:bpsR}
\tilde{\chi}^{R}_0 + \tilde{\chi}^{R}_{1\over 2}= q^{-n}\chi^{R}_{\left(0,{1\over 2}\right ),{1\over 2} + n}
\ee
and
\be
\tilde{\chi}^{R}_{1\over 2} + \tilde{\chi}^{R}_{4}= q^{-n}\chi^{R}_{\left ({1\over 2},{4}\right ),{1\over 2} + n}.
\ee
can give us the remaining two massless characters. In this sector, the BPS relations are essentially forced upon us as we approach the threshold weight, since we demand that the unitary massive state decomposes into (unitary) representations of the internal Ising subalgebra. If we then apply the spectral flow operator (i.e. Ising fusion rules) to the resulting Ramond ground states on the left-hand side of (\ref{eq:bpsR}) and write the result at the level of characters, we precisely reproduce the left-hand side of (\ref{eq:bps}) in the NS sector. Thus, since we have demonstrated that Ising fusion maps both the left and right hand sides of (\ref{eq:bps}) to those of (\ref{eq:bpsR}), and moreover that the relations (\ref{eq:bpsR}) are correct, we may feel reassured that employing (\ref{eq:bps}) and  (\ref{eq:bpsR}) in deriving our characters is justified.

One can do a similar analysis to that of the previous section to understand the modular properties of our conjectural Ramond sector characters. In the case of $\tilde\chi_0^R$, defining
\be
\overline\mu^R=f^{(5)}_{{\tau\over 2} + {1\over 2}}\left(6\tau,{\tau\over 2}\right)-f^{(5)}_{{\tau\over 2} + {1\over 2}}\left(6\tau,-{\tau\over 2}\right)+q^5f^{(5)}_{{\tau\over 2} + {1\over 2}}\left(6\tau,-{5\tau\over 2}\right)-q^5f^{(5)}_{{\tau\over 2} + {1\over 2}}\left(6\tau,{5\tau\over 2}\right),
\ee
this can be completed to a non-holomorphic weight $1$ modular form which transforms under $\Gamma_{0}(2)$,
\be
\hat{\overline \mu}^{R}(\tau,\overline \tau)=\overline\mu^{R}(\tau)-{1\over 2}\frac{1}{\sqrt{60 i}}\int_{-\overline \tau}^{i\infty} d\tau'~(\tau' + \tau)^{-{1\over 2}}\ubar{\theta}_{R}(\t) \cdot \ubar S(\tau')
\ee
where now
\[ \nonumber\ubar{\theta}_{R}(\t)=\left( \begin{array}{c}
\Theta^R_{(0, 1/2)} \\
\Theta^R_{(1/2, 4)}  \end{array} \right),\]
and $\ubar S(\tau)$ is the same as in the previous section. The factor $\frac{\eta(2\tau)^2}{\eta(\tau)^4}$ is a weight $-1$ modular form under $\Gamma_0(2)$, and thus $\tilde\chi_0^R$ as a whole transforms as a weight $0$ mock modular form under $\Gamma_0(2)$. $Z_{R, +}$ is a trace with periodic boundary conditions on the fermions on the space-like cycle of the torus and antiperiodic boundary conditions on the time-like cycle. It is straightforward to check that $\Gamma_0(2)$ is the congruence subgroup of $SL(2, \mathbb{Z})$ that preserves this spin structure. In particular, the spin structure is invariant under $\tau \mapsto \tau+1$ but not $\tau \mapsto -{ 1 \over \tau}$.

It's easy to see that under $T$ the $\Theta^R$ transform as
\be
\Theta^R_{(0, 1/2)}(\t+1) = e\left ({1\over 120}\right )\Theta^R_{(0, 1/2)}(\t),\qquad\text{and}\qquad\Theta^R_{(1/2, 4)}(\t+1) = e\left ({49\over 120}\right ) \Theta^R_{(1/2, 4)}(\t).
\ee
The second generator of $\Gamma_0(2)$ can be written as $ST^{-2}ST^{-1}$.

We can write the two generators of $\Gamma_0(2)$ as matrices which act on this vector in the following way (stripping off the overall factor of $\tau$ by convention):
\be
\rho(T)= \left( \begin{array}{cc}
e\left ({1\over 120}\right )& 0  \\
0 & e\left ({49\over 120}\right )  \end{array} \right)
\ee
and
\be
\rho(\tilde S)={i} \left( \begin{array}{cc}
e\left (-{15\over 120}\right )\sqrt{{2 \over (5- \sqrt{5})}} & e\left (-{39\over 120}\right )\sqrt{{2 \over (5+ \sqrt{5})}}  \\
e\left ({9\over 120}\right )\sqrt{{2 \over (5+ \sqrt{5})}} & e\left ({45\over 120}\right )\sqrt{{2 \over (5- \sqrt{5})}}  \end{array} \right)
\ee
where $\tilde S= ST^{-2}ST^{-1}$.

\section{Decomposition of the elliptic genus into characters}
\label{sec:decomps}

In this section we explicitly decompose the elliptic genera into our Spin(7) characters. We stress that although our closed-form expressions for the characters are conjectural, the decompositions are performed to finite order in $q$, where our numerics explicitly confirm the character formulae.

Recall our BPS equations, (\ref{eq:bps}). We will use these relations to write the decomposition of the elliptic genus in a suggestive way, completely analogous to the decompositions presented in \cite{Cheng:2014owa}.

Recall the NS-sector elliptic genus for a Spin(7) manifold $X$ is given by the trace:
\be
Z_{NS,+}(\tau)=\Tr_{NS,R} (-1)^{F_R} q^{L_0- \frac{c}{24}}\overline{q}^{\overline{L_0} - \frac{c}{24}}.
\ee
As discussed in \S\ref{sec:genera}, up to a constant term determined by the Euler character of $X$ (which vanishes
if $\chi(X) = 24$), 
this has the $q$-expansion
\be
Z_{NS,+} = {1\over \sqrt q} + 276 \sqrt q + 2048 q + 11202 q^{3/2} + \ldots
\ee
In what follows we simply set the constant term to zero. We discuss the possible meanings of the decompositions in this section in the conclusion; and in the companion \cite{M5spin7} we describe
some precise realizations of super-modules with vanishing constant which realize the various
symmetries we discuss below.

Firstly, we note that these coefficients decompose nicely into dimensions of irreducible representations of Co$_1$. For example, 276 is itself a dimension of an irrep of Co$_1$. Additionally, $2048 = {\bf 1} + {\bf 276} + {\bf 1771},  11202 = {\bf 1} + {\bf 276} + {\bf 299} + {\bf 1771} + {\bf 8855}$, etc.  In fact the decomposition into ${\cal N}=1$ super-Virasoro characters would suggest a relation to the Conway module described in \cite{JohnConway} and recently reviewed in \cite{Cheng:2014owa}. 

Instead, we would like to decompose into Spin(7) characters. The most general decomposition into Ramond sector characters of the Spin(7) algebra has the form
\be
Z_{R, +}= A_0 \tilde{\chi}_0^R + A_{1\over 2} \tilde{\chi}_{1\over 2}^R + A_{4}\tilde{\chi}_{4}^R + \sum_{n=1}^\infty B_n \chi_{\left(0,{1\over 2}\right ),{1 \over 2} + n}^R + \sum_{n=1}^\infty C_n\chi_{\left ({1\over 2},{4}\right ),{1\over 2} + n}^R
\ee
where now $A_0, A_{1\over2}, A_{4}, B_n$, and $C_n$ are all (positive integer) constants. The NS$, +$ elliptic genus has a similar expansion in terms of the NS characters, with coefficients $a_0, a_{1/2}, a_{4}, b_n, c_n$. The isomorphism between Ramond and NS sectors give us the following relations between expansion coefficients: $a_0=A_{4}$, $a_{1\over2}=A_{1\over2}$, $a_{4}=A_0$, $b_n=C_n$, and $c_n=B_n$. 

Now we have enough information to fix all of the constants in the decomposition. Using that $a_0=A_{4}=1$ and $a_{1\over2}=A_{1\over2}=0$, at O(1) in the decomposition of $Z_{R, +}$ we get that $A_0=a_{4}=23$. With this coefficient fixed, the entire decomposition is now fixed.

Once these coefficients are fixed, we get the following for the first few massive coefficients,
\be
b_1=253,\qquad b_2=7359,\qquad  b_3=95128, \ldots
\ee
and
\be
c_1=1771,\qquad  c_2=35650,\qquad c_3=374141, \ldots
\ee 
Using the above relations, we can repackage the decomposition in the following form, which is entirely analogous to the $\mathcal N=2$ and $\mathcal N=4$ decompositions studied in \cite{Cheng:2014owa}. First define

\begin{align}\label{fone}
\nonumber f_1(\tau)&=q^{-{1\over 120}}(-1 + c_1 q + c_2 q^2 + c_3 q^3 + \ldots) \\
&=q^{-{1\over 120}}(-1 + 1771 q + 35650 q^2 + 374141 q^3 + \ldots)
\end{align}
and 
\begin{align}\label{fseven}
\nonumber f_7(\tau)&= q^{-{49\over 120}}(1 + b_1 q + b_2 q^2 + b_3 q^3 + \ldots),\\
&=q^{-{49\over 120}}(1 + 253 q + 7359 q^2 + 95128 q^3 + \ldots).
\end{align}
Then we can rewrite
\be\label{eq:ZR}
Z_{R, +}= 2\frac{\eta(2\tau)^2}{\eta(\tau)^4}\left (24 \overline\mu^R +  f_1(\tau)\Theta^R_{(0, 1/2)}(\t)+f_7(\tau) \Theta^R_{(1/2, 4)}(\t) \right)
\ee
or, in the NS sector,
\be\label{eq:ZNS}
Z_{NS, +}= \frac{\eta(\tau)^2}{\eta\left ({\tau\over 2}\right)^2 \eta(2\tau)^2}\left (24 \overline\mu^{NS} + f_1(\tau) \Theta^{NS}_{1/2}(\t) + f_7(\tau)\Theta^{NS}_{0}(\t)\right),
\ee
where $f_1(\tau)$ and $f_7(\tau)$ are as defined above, and $\overline\mu^{NS}$ and $\overline \mu^R$ are as defined in the previous section. Using  the transformation properties of these functions derived in the previous section, we can predict how the functions $f_1(\t)$ and $f_7(\t)$ transform under $SL(2,\mathbb Z)$.

In particular, the modular group $SL(2,\mathbb Z)$ is generated by the matrices $T=\sm1101$, and ${S=\sm01{-1}0}$.  The multiplier system $\rho$ of $\ubar f$ on the entirety of  $SL(2,\mathbb Z)$ is generated by
\begin{align}
\rho(T) &=   \left( \begin{array}{cc}
e\left (-{1\over 120}\right )& 0  \\
0 & e\left (-{49\over 120}\right )  \end{array} \right)&\rho(S) =   e(\tfrac18)\left(
\begin{array}{cc}
 -\sqrt{\frac{2}{5+\sqrt{5}}} & \sqrt{\frac{2}{5-\sqrt{5}}} \\
 \sqrt{\frac{2}{5-\sqrt{5}}} & \sqrt{\frac{2}{5+\sqrt{5}}} \\
\end{array}
\right).
\end{align}
Mock modularity and Rademacher summability have been proposed as a hallmark of moonshine~\cite{Cheng:2011tl}.  Using standard techniques for computing vector-valued Rademacher sums\cite{Whalen:2014vq}, it can be seen that the weight-$\frac12$ Rademacher sum generated by the polar part $(q^{-\frac{1}{120}},q^{-\frac{49}{120}})$ and above multiplier system over $SL(2,\mathbb Z)$ gives $\ubar f$ exactly.  The Rademacher summability of the $\ubar f$ serves as a weak independent verification of the decomposition.

It will strike the attentive reader in the post-EOT era that the integer coefficients appearing in the $q$-series
(\ref{fone}) and (\ref{fseven}) are related in a simple way to irreducible representations of sporadic simple groups.  For
instance, in $f_7$, one can decompose into dimensions of irreps of M$_{24}$ as
$$253 = \textbf{253}, ~~7359 = \textbf{23} \oplus \textbf{252} \oplus \textbf{770} \oplus {\overline {\textbf{770}}} \oplus \textbf{5544}$$
$$95128 = \textbf{253} \oplus \textbf{990} \oplus {\overline {\textbf{990}}} \oplus 2 \times \textbf{1265}  \oplus \textbf{1771} \oplus \textbf{2024} \oplus \textbf{2277} \oplus \textbf{3312} $$
$$\oplus 2 \times \textbf{3520} \oplus 2 \times \textbf{5313} \oplus \textbf{5544} \oplus \textbf{5796}  \oplus  5 \times \textbf{10395},
\ldots 
$$
\noindent
while in $f_1$, one sees
$$1771 = \textbf{1771}, ~~35650 = \textbf{252} \oplus \textbf{253} \oplus \textbf{1771} \oplus 2 \times \textbf{3520} \oplus \textbf{5544} \oplus 2 \times \textbf{10395}, \ldots$$
\noindent
The coefficients grow quickly enough that one needs some confidence in the actual existence of
an M$_{24}$ symmetry, and a systematic means of determining compositions, to go further.  
In a companion paper \cite{M5spin7}, we describe how in a precise chiral CFT with the same chiral algebra and giving rise to the same (chiral) partition function, we
can realize an M$_{24}$ symmetry with explicit decompositions (and checks of twining for all M$_{24}$ 
conjugacy classes) at all orders.  This verifies (in a specific example) that the numerology above
hides deeper meaning in some specific instance.  

Similar decompositions into representations of the sporadic groups Co$_2$ and Co$_3$ exist and
will be elaborated upon in \cite{M5spin7}.

\section{Conclusions}
\label{sec:conclusions}

In this paper, we studied the elliptic genera of manifolds of exceptional holonomy Spin(7).  These
constitute one of the few classes of possible supersymmetric compactification manifolds for
superstrings.

In order to look for hidden  connections between the geometry of these spaces and number theory and
representation theory, following the example of \cite{EOT}, it is natural to decompose the
elliptic genus into characters of the relevant chiral algebra (${\cal SW}(3/2,2)$ at $c=12$), and
try to find representation-theoretic interpretations for the coefficients.
While we find suggestive results in \S\ref{sec:decomps}, we note that a precise relationship of any particular
Spin(7) manifold with the group M$_{24}$ (or the groups Co$_2$ and Co$_3$, which also appear to
play a special role in the $q$-expansions of the functions $f_1(\tau), f_7(\tau)$) is not indicated
by our results. 
For a given explicit manifold $X$, for generic $\chi(X)$ there would be an unwanted constant term in the
elliptic genus.  For those spaces which have $\chi(X) = 24$ and vanishing constant term, one would expect the space
to at most admit discrete symmetries whose twinings could agree with M$_{24}$ for a handful of
conjugacy classes (of the same orders).  Even in the case of the K3 conformal field theory, where (obviously) a unique candidate topology exists realizing the suggestive elliptic genus, a precise statement of
Mathieu moonshine has been elusive.

However, there is a simple explicit chiral conformal field theory with the same chiral algebra
we studied here,
which manifestly realizes the M$_{24}$ moonshine hinted at in \S\ref{sec:decomps}.  That is the subject of the
companion paper \cite{M5spin7}.  One could hope that this chiral CFT is a sort of ``Platonic model,"
indicating the largest symmetries that might be realized at (perhaps non-geometric) points in 
the moduli space of Spin(7) compactifications.  Broadly similar remarks apply to the
constructions based on the  ${\cal N}=2, 4$ superconformal algebras
in the recent paper \cite{Cheng:2014owa}, and compactifications on Calabi-Yau
4-folds or 8-dimensional hyperK\"ahler spaces.

\medskip
\centerline{\bf{Acknowledgements}}
\medskip

We would like to thank V. Buciumas, D. Bump, M. Cheng, X. Dong, J. Duncan, D. Ramirez, and
T. Wrase for helpful discussions.  We especially thank Ryan North of Dinosaur Comics$^{\text{TM}}$ for
permission to use awesome graphics.  N.B. is supported by a Stanford Graduate Fellowship and N.M.P. is supported
by an NSF Graduate Research Fellowship.  S.M.H. is supported by the Harvard University Lawrence Golub Fellowship in the Physical Sciences.  S.K. and D.P.W. acknowledge the support of the NSF
via grant PHY-0756174, DOE Office of Basic Energy Sciences contract DE-AC02-76SF00515, and
the John Templeton Foundation.  

\appendix
\section{Jacobi Theta Functions and the Dedekind Eta Function} \label{app:a}

The Jacobi theta functions are a collection of weight $\half$, index 1 Jacobi forms.  Write $q=\exp(2\pi i\tau)$ and $y=\exp(2\pi i z)$.  Then the Jacobi theta functions can be written as 
\begin{align}
\theta_1(\tau,z) 
&= -i q^{\frac{1}{8}} y^{\half} \prod_{n=1}^{\infty} (1-q^n)(1-y q^n)(1 - y^{-1} q^{n-1})  
= i \sum_{n=-\infty}^{\infty} (-1)^n q^{\frac{(n-\half)^2}{2}} y^{n-\half} \nn \\
\theta_2(\tau,z) 
&=q^{\frac{1}{8}} y^{\half} \prod_{n=1}^{\infty} (1-q^n) (1+y q^n) (1 + y^{-1} q^{n-1})  
= \sum_{n=-\infty}^{\infty} q^{\frac{(n-\half)^2}{2}} y^{n-\half} \nn \\
\theta_3(\tau,z) 
&= \prod_{n=1}^{\infty} (1-q^n) (1 + y q^{n-\half}) (1 + y^{-1} q^{n-\half})  
= \sum_{n=-\infty}^{\infty} q^{\frac{n^2}{2}} y^n \nn \\
\theta_4(\tau,z)
&= \prod_{n=1}^{\infty} (1-q^n) (1 - y q^{n-\half}) (1 - y^{-1} q^{n-\half}) 
= \sum_{n=-\infty}^{\infty} (-1)^n q^{\frac{n^2}{2}} y^n.
\end{align}

The Dedekind Eta function is a modular function of weight $\half$ which is defined to be
\be
\eta(\tau) = q^{\frac{1}{24}} \prod_{n=1}^{\infty} (1-q^n)=q^\frac{1}{24}\sum_{n=-\infty}^\infty (-1)^n q^\frac{3n^2-n}{2}.
\ee

\section{$\mathcal{SW}(3/2, 2)$ Algebra}
\label{ap:comm}

Here we set our notation for the algebra $\mathcal{SW}(3/2, 2)$ at $c=12$ and write down the commutation relations. The $\mathcal{SW}(3/2, 2)$ algebra has a basis that consists of operators $L_n, X_n,$ for $n\in\bbZ$ and $M_m, G_m$ for $m\in\bbZ+\frac{1}{2}$.  We define $L$ and $X$ to be bosonic operators, and $M$ and $G$ to be fermionic.  

The normal-ordered product is given by
\begin{equation}\label{nop}
\lc XY\rc_n= \sum_{k\leq -\Delta_{X}}X_kY_{n-k}+(-1)^{F}\sum_{k>-\Delta_{X}}Y_{n-k}X_{k}.
\end{equation}
where $(-1)^{F}$ is $-1$ if both $X$ and $Y$ are fermionic and 1 otherwise.  The $\Delta$s are the conformal weights of the operators.  Recall that $\Delta_L=\Delta_X=2$, $\Delta_G=\frac{3}{2}$ and $\Delta_M = \frac{5}{2}$.

We are interested in constructing unitary irreducible highest weight representations, which are generated by the action of the algebra on a highest weight state $|h, x\rangle$, where $h$ and $x$ are the eigenvalues of $L_0$ and $X_0$ respectively.

We can construct a Hermitian conjugate on this space.
The operators in the algebra are also given Hermitian conjugates: $L^\dagger_n = L_{-n}$, $X^\dagger_n = X_{-n}$, and $G^\dagger_n = G_{-n}$.  The $M$ operator possesses a non-standard conjugate: ${M^\dagger_n=-M_{-n}-\frac{1}{2}G_{-n}}$. 

The commutation relations of the algebra are given below\footnote{We correct some small typos on p. 35 of \cite{shatashvili1995superstrings}.}:
\begin{align}
[L_n,L_m]&= (n - m) L_{n + m} + (n^3 - n) \delta_{m+n} \nn \\
[X_n,X_m]&= 
 \tfrac83 (n^3 - n) \delta_{n+m} + 
  8 (n - m) X_{n + m}\nn \\
[L_n, X_m]&= 
 \tfrac13 (n^3 - n) \delta_{n+m} + (n - m) X_{n + m}\nn\\
[G_n, X_m]&= 
 \thalf \left(n + \thalf\right) G_{n+m} + M_{n+m}\nn\\
[X_n, M_m]&= \left(\tfrac{15}4 \left(n + 1\right) \left(m + \tfrac32\right) - 
     \tfrac54 \left(n + m + \tfrac32\right) \left(n + m + \tfrac52\right)\right) G_{n + m}\nn\\&\qquad - \left(- 8 \left(n + 1\right) + \tfrac{11}2 \left(n + m + \tfrac52\right)\right) M_{n + m} - 6\lc GX\rc_{n + m}\nn\\
[L_n, G_m]&= \left(\thalf n - m\right) G_{n+m}\nn\\
[L_n, M_m]&= 
 \qrt n (n+1) G_{n+m} + \left( \tfrac32 n-m\right) M_{n+m}\nn\\
\{G_n, M_m\}&= 
 \tfrac23 \left(n^2 - \qrt\right) \left(n - \tfrac32\right) \delta_{n+m} - \left(n + \thalf\right) L_{n+m} + (3 n - m) X_{n+m}\nn\\
\{M_n, M_m\}&= -\tfrac83 \left(n^2 - \tfrac94\right) \left(n^2 - \qrt\right) \delta_{n+m}+
  \Big(\tfrac{15}2 \left(m + \tfrac32\right) \left(n + \tfrac32\right)\nn\\&\qquad - \tfrac52 \left(n + m + 2\right) \left(n + m + 3\right)\Big) L_{n+m}+
  \Big(16 \left(m + \tfrac32\right) \left(n + \tfrac32\right)\nn\\&\qquad - \tfrac52 \left(n + m + 2\right) \left(n + m + 3\right)\Big) X_{n + m} -  12 \lc LX\rc_{ n + m} +  
  6 \lc GM\rc_{n + m}\nn\\
\{G_n, G_m\}&= 
 (4n^2 - 1) \delta_{n+m}+ 2 L_{n+m}.
\end{align}

\section{Subsingular vectors}\label{app:subsingular}

For the sake of completeness and to assist in replicating our results, we provide a construction of a subsingular vector from the module generated by $|\half,\half\rangle$.  The Verma module possess a subsingular vector with $L_0$ eigenvalue of 4 and $X_0$ eigenvalue of 20.  Up to addition of an element from the subspace generated by singular vectors, the subsingular vector is at

\begin{align*}&\Big(G_{-7/2} + \tfrac{7}{3} G_{-5/2} X_{-1} - 
 \tfrac{2}{3} M_{-5/2}X_{-1} + 2 X_{-3} G_{-1/2} + 
 \tfrac{10}{9} X_{-2} G_{-3/2} + \tfrac{8}{9} X_{-2} M_{-3/2}\\&\phantom{aaa}+ 
 \tfrac{1}{3} G_{-3/2}(X_{-1})^2 + 
 \tfrac{5}{3} X_{-2}X_{-1} G_{-1/2} + 
 \tfrac{1}{6} X_{-2}X_{-1}M_{-1/2} + 
\tfrac{23}{21} L_{-1}(X_{-1})^{2} M_{-1/2}\\&\phantom{aaa}- 
 \tfrac{11}{192} (X_{-1})^3 G_{-1/2} - 
\tfrac{33}{224} (X_{-1})^3 M_{-1/2}\Big) \ket{\thalf, \thalf}.
 \end{align*}
 
The $|h=0, x=0\rangle$ Verma module possess a subsingular vector with $L_0$ eigenvalue of 11/2 and $X_0$ eigenvalue of 65/2, but the subsingular vector is too complex to usefully write down.

\label{referenceslol}
\bibliographystyle{JHEP}
\bibliography{refs}
    \end{document}